\documentclass[10pt,twocolumn,showpacs,showkeys,preprintnumbers,amssymb,aps,superscriptaddress,pre]{revtex4-2}
\usepackage{amsmath, amssymb}
\usepackage{graphicx}
\usepackage{color,array}
\usepackage[colorlinks={true}]{hyperref}
\hypersetup{colorlinks=true,linkcolor=red,citecolor=blue,urlcolor=blue}
\usepackage{bm}
\usepackage{amsmath}
\usepackage{appendix}
\usepackage{comment}
\usepackage{orcidlink}

\begin{document}

\preprint{APS/123-QED}

\title{Emergent Wigner magnon crystals in a fully frustrated Heisenberg four-leg tube}

\author{Azam Zoshki\,\orcidlink{0000-0002-0023-6094}}
\author{Hamid Arian Zad\,\orcidlink{0000-0002-1348-1777}}
\author{Jozef Stre\v{c}ka\,\orcidlink{0000-0003-1667-6841}}
\email{Corresponding author: jozef.strecka@upjs.sk}
\affiliation{Department of Theoretical Physics and Astrophysics, Faculty of Science, 
P. J. \v{S}af\'arik University, Park Angelinum 9, 041 54 Ko\v{s}ice, Slovak Republic}

\date{\today}

\begin{abstract}
The ground state, magnetization curves, low-temperature thermodynamics, and heat-engine performance of the fully frustrated spin-$1/2$ Heisenberg four-leg tube with antiferromagnetic inter- and intra-plaquette coupling constants $J_1$ and $J_2$ are examined using exact diagonalization, density matrix renormalization group, and localized-magnon theory. In the unfrustrated to weakly frustrated regime $J_2/J_1 \ll 2$, the system exhibits a continuous field-driven quantum phase transition between the gapped Haldane phase and a gapless Tomonaga-Luttinger quantum spin liquid. In the highly frustrated regime $J_2/J_1 \geq 2$, the system contrarily displays discontinuous field-driven quantum phase transitions between the Wigner magnon crystals, which are manifested in zero-temperature magnetization curves as intermediate plateaus at zero, one-quarter, one-half, and three-quarters of the saturation magnetization. The low-temperature magnetic and thermodynamic properties in this regime are accurately captured by an effective interacting lattice-gas model of two monomer quasi-particle species constructed from localized one- and two-magnon states. Finally, we explore a quantum Stirling heat engine using the fully frustrated four-leg tube as the working medium. The work output and efficiency are strongly suppressed near discontinuous field-induced transitions and reach pronounced local maxima well inside the stability regions of the Wigner magnon crystals.
\end{abstract}
 
\maketitle

\section{Introduction} \label{sec:Introduction}

Geometric spin frustration \cite{Wannier1950,Anderson1956,Anderson1973,Ramirez1991} is an essential ingredient for detecting various unconventional collective phenomena in quantum Heisenberg antiferromagnets \cite{Furukawa2015,Honecker2016,Ramirez2025}. A mutual interplay of strong quantum fluctuations, geometric spin frustration, and external magnetic field often stabilizes exotic quantum magnetic orders \cite{Shas,Jian,Richter}, drive quantum phase transitions and criticality \cite{Koga,Kim,Gane}, and/or generate exotic quantum spin liquids \cite{Sava,Bro,Liu}. Quantum spin ladders constitute an important intermediate geometry between one- and two-dimensional systems, while imposing periodic boundary conditions in the transverse direction transforms an $n$-leg ladder ($n \geq 3$) into a spin tube and may introduce additional frustration. This is particularly evident for tubes with an odd number of legs, where antiferromagnetic interactions around the transverse rings are intrinsically frustrated, giving rise to gapped dimerized ground states, chiral degrees of freedom, and fractional magnetization plateaus \cite{Schnack2004,Plat2012,Tachibana2020}. Even-leg spin tubes are not geometrically frustrated by nearest-neighbor interactions alone, but frustration can be generated by additional competing exchange interactions, opening the possibility of similarly rich quantum behavior \cite{Garlea2008, Zheludev2008, Arlego2011, Arlego2013, GomezAlbarracin2014, Ramos2014,Jafari2019}.

An important experimental motivation for frustrated four-leg spin tubes is provided by the magnetic compound $\mathrm{Cu}_2\mathrm{Cl}_4 \cdot \mathrm{D}_8\mathrm{C}_4\mathrm{SO}_2$ \cite{Garlea2008}, which was identified by inelastic neutron scattering as an almost ideal experimental realization of the quasi-one-dimensional spin-$1/2$ Heisenberg four-leg tube \cite{Zheludev2008}. Its dominant antiferromagnetic exchange interactions form four coupled spin chains, while additional competing interactions diagonally coupling adjacent legs introduce geometric spin frustration and stabilize slightly incommensurate spin correlations. This experimental realization stimulated theoretical interest in frustrated Heisenberg four-leg spin tubes, in which competing inter-plaquette exchange interactions give rise to several unconventional gapped quantum phases such as plaquette, incommensurate and Haldane phases alongside with the gapless Tomonaga-Luttinger quantum spin liquid \cite{Arlego2011,Arlego2013}. More recently, Jafari \textit{et al.} \cite{Jafari2019} reported for the frustrated Heisenberg four-leg tube in the strong leg-coupling regime both continuous and discontinuous field-induced phase transitions closely associated with the presence of fractional magnetization plateaus.  

The ground-state phase diagram of the frustrated spin-$1/2$ Heisenberg four-leg tube in an external magnetic field was particularly investigated by Gomez Albarrac\'in \textit{et al.} \cite{GomezAlbarracin2014}. In the regime of weakly interacting square plaquettes, they demonstrated that low-energy singlet, triplet, and quintuplet plaquette states play a crucial role in the formation of fractional magnetization plateaus. These results suggest that the field-induced phases of frustrated Heisenberg four-leg tubes can be naturally understood in terms of specific low-energy states of the elementary square plaquettes. A particularly intriguing situation emerges in the fully frustrated limit considered in the present work, where the special geometry of the inter-plaquette interactions enables an exact localization of one- and two-magnon states on individual square plaquettes. The term fully frustrated refers to all-to-all connection between spins of adjacent square plaquettes and this particular lattice geometry was studied in detail by Honecker \textit{et al.} \cite{Honecker2000} with the special emphasis laid on two- and three-leg ladders. Although an experimental realization of such a fully connected system has yet to be identified, its rich behavior arising from the interplay of high symmetry and geometric frustration provides valuable insight into fundamental aspects of quantum magnetism of highly frustrated quantum magnets.

Presence of localized magnons \cite{Schmi2022, Zoghlin03} generally provides a powerful framework for understanding unconventional field-induced phases of highly frustrated quantum magnets \cite{Derzhko2006,Derzhko2015}. Owing to destructive quantum interference, a magnon may become strictly localized within a finite trapping cell, giving rise to dispersionless one-magnon bands and exact many-magnon eigenstates. At sufficiently high magnetic fields, independent localized magnons can form spatially ordered states often referred to as Wigner magnon crystals, which manifest themselves through magnetization plateaus and jumps and characteristic low-temperature thermodynamic behavior. Moreover, the restricted low-energy manifold of localized-magnon states can frequently be mapped onto an effective classical lattice-gas model, providing access to thermodynamic properties in the vicinity of the corresponding field-induced transitions \cite{Derzhko2006,Derzhko2015}. In the fully frustrated Heisenberg four-leg tube considered here, this concept can be extended beyond standard localized-magnon theories usually valid only at high magnetic fields near the saturation, since the elementary square plaquettes can host both bound one- and two-magnon states. Their field-controlled spatial ordering consequently gives rise to several Wigner magnon crystals stabilized over the entire field range from zero up to saturation field.

In the present work, we investigate the ground-state and finite-temperature properties of the fully frustrated spin-$1/2$ Heisenberg four-leg tube in an external magnetic field over a wide range of the frustration ratio $J_2/J_1$. By combining density-matrix renormalization group and exact-diagonalization calculations with localized-magnon theory, we construct a comprehensive ground-state phase diagram and magnetization process. We further construct an effective interacting lattice-gas model of the relevant localized-magnon states and demonstrate that it accurately captures the low-temperature magnetothermodynamic properties of the highly frustrated four-leg tube. The pronounced field dependence of the low-temperature thermodynamic properties further motivates us to explore the fully frustrated four-leg tube as a working medium of a quantum Stirling heat engine \cite{Yin,Purk,Cruz2023,Wang2024}. We analyze a magnetic Stirling cycle composed of two isothermal and two isofield processes and determine its heat exchange, work output, and efficiency. 

The remainder of the paper is organized as follows. In Sec.~\ref{sec:Model}, we introduce the fully frustrated spin-$1/2$ Heisenberg four-leg tube and briefly describe the numerical and analytical methods employed. Section~\ref{sec:GSPD} presents the ground-state phase diagram and magnetization process. In Sec.~\ref{sec:magnetothermodynamics}, we develop the effective lattice-gas description and analyze the finite-temperature magnetothermodynamic properties in the highly frustrated regime. Section~\ref{sec:heatengine} is devoted to the performance of the quantum Stirling heat engine using the four-leg tube as a working medium. Finally, Sec.~\ref{sec:conclusions} summarizes our main results and conclusions.

\section{Model and Method}
\label{sec:Model}

\begin{figure}
\centering
\includegraphics[width=0.45\textwidth]{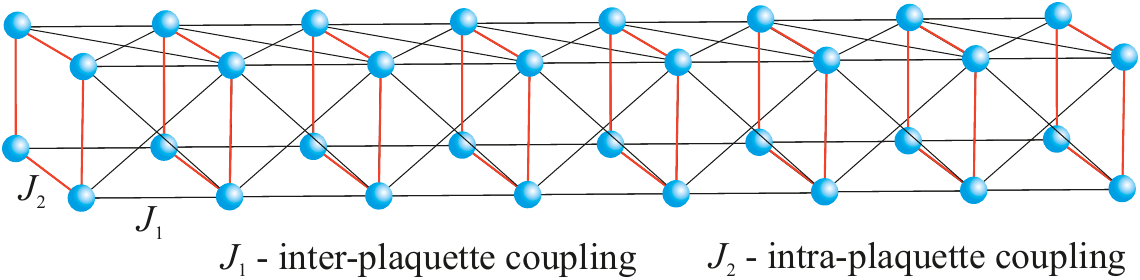}
\vspace{-0.3cm}
\caption{Schematic illustration of a fully frustrated four-leg tube. The blue balls represent lattice position of spin-1/2 particles, thick red lines indicate intra-plaquette interaction $J_2$, and thin black lines represent inter-plaquette interaction $J_1$. For better clarity, the diagonal inter-plaquette couplings $J_1$ are shown only for top and front square plaquettes.}
	\label{fig:4LegL_model}
\end{figure}

We consider a fully frustrated spin-1/2 Heisenberg four-leg tube, which can alternatively be viewed as a one-dimensional array of interacting square plaquettes schematically illustrated in Fig.~\ref{fig:4LegL_model}. 
The Hamiltonian of this one-dimensional frustrated quantum spin model is given by the following expression:
\begin{eqnarray}\label{Eq:4ll_ham}
	\hat{H} &=& J_1 \sum_{i=1}^{N_u} \Biggl(\sum_{j=1}^{4}	\hat{\mathbf{S}}_{i,j} \Biggr) \!\cdot\! \Biggl(\sum_{k=1}^{4}	\hat{\mathbf{S}}_{i+1,k} \Biggr)
	\nonumber\\
	&+& J_2 \sum_{i=1}^{N_u} \Big(
	\hat{\mathbf{S}}_{i,1} \!\cdot\! \hat{\mathbf{S}}_{i,2}
	+ \hat{\mathbf{S}}_{i,2} \!\cdot\! \hat{\mathbf{S}}_{i,3}
	+ \hat{\mathbf{S}}_{i,3} \!\cdot\! \hat{\mathbf{S}}_{i,4}
	+ \hat{\mathbf{S}}_{i,4} \!\cdot\! \hat{\mathbf{S}}_{i,1}
	\Big) \nonumber\\
	&-& h \sum_{i=1}^{N_u} \left(\hat{S}_{i,1}^z + \hat{S}_{i,2}^z + \hat{S}_{i,3}^z + \hat{S}_{i,4}^z\right),
\end{eqnarray}
where $\hat{\boldsymbol S}_{i,j}=(\hat{S}_{i,j}^x,\hat{S}_{i,j}^y,\hat{S}_{i,j}^z)$ denotes the spin-1/2 operator at the lattice site specified by the indices $i$ and $j$, the former index $i=1,\ldots,N_u$ labels the unit cell (square plaquette), while the latter index $j=1,\ldots,4$ specifies one of the four spins belonging to the $i$-th square plaquette ($N_u$ denotes the total number of unit cells). The antiferromagnetic coupling constants $J_1>0$ and $J_2>0$ determine the inter-plaquette and intra-plaquette interactions, respectively. The last term $h=g \mu_\text{B} B$ introduces Zeeman's energy associated with an external magnetic field $B$, $g$ is the Land\'e g-factor, and $\mu_\text{B}$ is the Bohr magneton. To eliminate the finite-size effects, we impose the periodic boundary condition $S_{N_u+1, j} \equiv S_{1, j}$ along the longitudinal direction of the tube. By utilizing a combination of complementary analytical and numerical techniques described in the subsequent parts, we aim to solve the model and investigate its physical properties.

\subsection{Bound one-magnon state and magnon crystals}
\begin{figure}
	\centering
	\resizebox{0.48\textwidth}{!}{
		\includegraphics[trim = 0 0 0 0, clip]{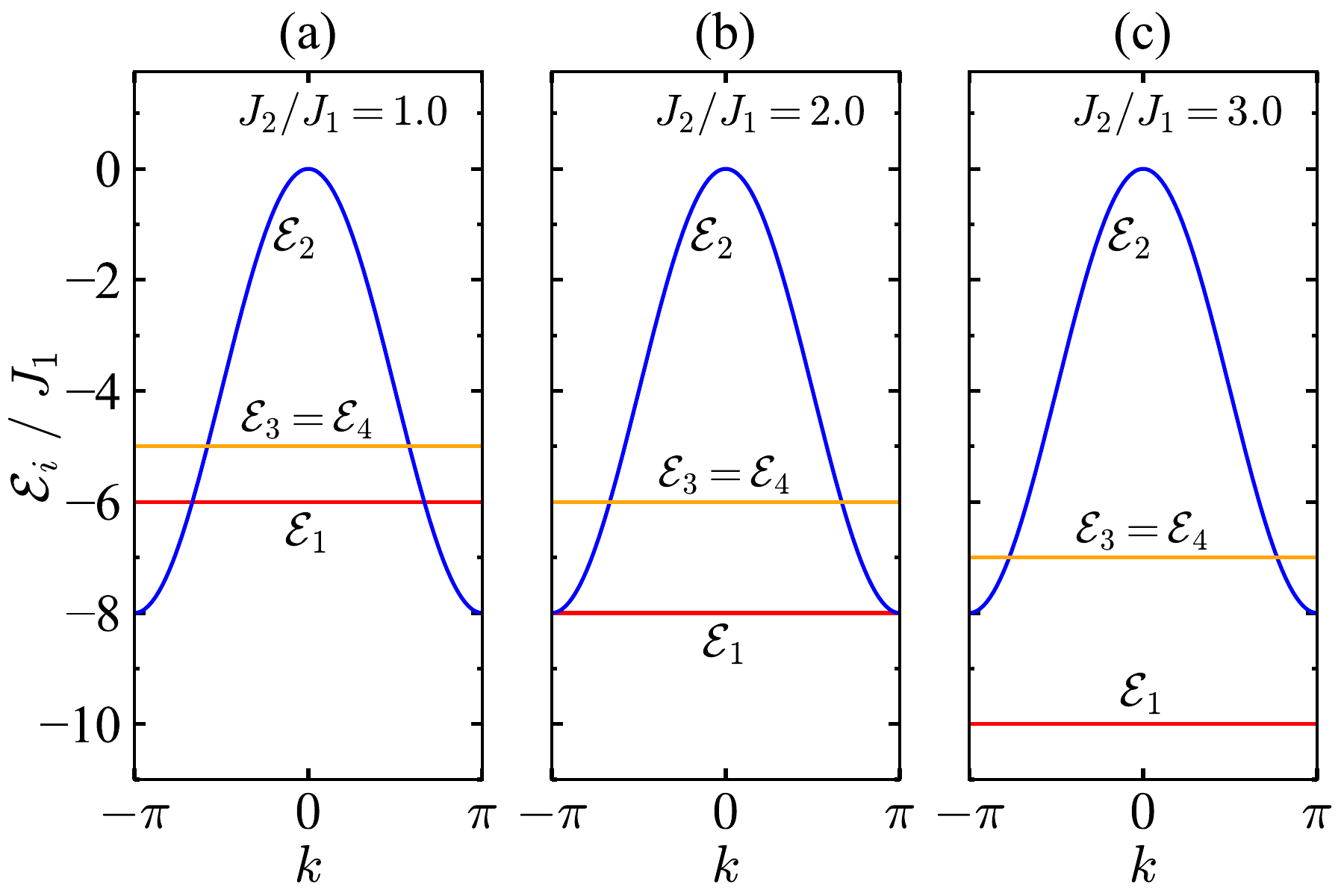}}
	\vspace{-0.7 cm}
	\caption{The one-magnon energy bands of the fully frustrated spin-1/2 Heisenberg four-leg tube given by Eq. (\ref{Eq:Epsilons}) at zero magnetic field for three selected values of the interaction ratio: (a) $J_2/J_1 = 1.0$; (b) $J_2/J_1 = 2.0$; (c) $J_2/J_1 = 3.0$.}
	\label{fig:flatband}
\end{figure}

To gain insight into eigenstates emerging just below the saturation field, we first exactly determine the one-magnon spectrum of the fully frustrated spin-1/2 Heisenberg four-leg tube. In the one-magnon sector with the $z$-component of the total spin $S^z_{tot} = 2N_u - 1$, the exact eigenstates can be constructed within the orthonormal basis $|i, j\rangle = \hat{S}_{i,j}^- |\text{FM}\rangle$ ($i \in \{1, \dots, N_u\}$, $j \in \{1, 2, 3, 4\}$). These basis states are obtained by applying a single spin-lowering operator $\hat{S}_{i,j}^-$ to the fully polarized ferromagnetic state: 
\begin{equation}\label{Eq:FM-State}
	\vert\text{FM}\rangle=\prod_{i=1}^{N_u}\vert \uparrow_{1,i} \uparrow_{2,i} 
	\uparrow_{3,i} \uparrow_{4,i} \rangle,
\end{equation}
which represents a trivial eigenstate of the Hamiltonian (\ref{Eq:4ll_ham}) with the energy $E_\text{FM} = N_u(4J_1 + 2J_2 - 2h)$. Applying the Hamiltonian (\ref{Eq:4ll_ham}) to this one-magnon basis leads to a system of coupled linear equations as detailed in Appendix \ref{app:derivation}. Exploiting translational invariance along the tube axis, the eigenvalue problem $\hat{H}|\Psi_k\rangle = E_k |\Psi_k\rangle$ can be solved using the Bloch ansatz $|\Psi_k\rangle = \sum_{i=1}^{N_u} \sum_{j=1}^{4} c_{j,k} {\rm e}^{{\rm i} k x_i} |i, j\rangle$, where $k \in [-\pi, \pi]$ is the wave number in the first Brillouin zone and $x_i$ denotes the position of the $i$-th unit cell along the tube axis. The one-magnon excitation energy given relative to the energy of the ferromagnetic eigenstate 
$\mathcal{E}_k = E_k - E_{\text{FM}}$ is then determined by the roots of the characteristic equation obtained from the following $4 \times 4$ secular determinant (see Appendix \ref{app:derivation}):
\begin{equation}
	\begin{vmatrix}
		\mathcal{A} & \mathcal{B} & J_1 \cos k & \mathcal{B} \\
		\mathcal{B} & \mathcal{A} & \mathcal{B} & J_1 \cos k \\
		J_1 \cos k & \mathcal{B} & \mathcal{A} & \mathcal{B} \\
		\mathcal{B} & J_1 \cos k & \mathcal{B} & \mathcal{A}
	\end{vmatrix}
	= 0.
\end{equation}
where $A = J_1(\cos k - 4) - J_2 + h -\mathcal{E}_k$ and $\mathcal{B}=J_1 \cos k + \frac{J_2}{2}$. Solving this characteristic equation yields the four branches of the one-magnon energy spectrum:
\begin{eqnarray}\label{Eq:Epsilons}
	\mathcal{E}_1 &=& -4J_1 - 2J_2 + h, \nonumber\\
	\mathcal{E}_2 &=& -4J_1 + 4J_1 \cos k + h, \nonumber\\
	\mathcal{E}_{3,4} &=& -4J_1 - J_2 + h. 
\end{eqnarray}
Among the four one-magnon branches (\ref{Eq:Epsilons}), three branches $\mathcal{E}_1$ and $\mathcal{E}_{3,4}$ are dispersionless flat bands associated with bound one-magnon eigenstates localized on individual square plaquettes. Fig. \ref{fig:flatband} shows the one-magnon bands $\varepsilon_1$ (red), $\varepsilon_2$ (blue), and $\varepsilon_3 = \varepsilon_4$ (orange) as a function of the wave number $k$ for three different values of the interaction ratio $J_2/J_1$. In contrast to the dispersive band $\varepsilon_2$, the three flat bands $\varepsilon_1$ and $\varepsilon_{3,4}$ shift to lower energies with increasing of the interaction ratio $J_2/J_1$. A comparison of the eigenenergies (\ref{Eq:Epsilons}) of the four one-magnon branches shows that the flat band with the energy $\mathcal{E}_1$ is the lowest-energy branch in the highly frustrated regime $J_2>2J_1$, whereas the dispersive band with the energy $\mathcal{E}_2$ becomes the lowest-energy branch in the reverse case. 
The bound one-magnon eigenstate associated with the flat band $\mathcal{E}_1$ has character of a localized triplet state on the $i$-th square plaquette:
\begin{eqnarray}
|\text{t}\rangle_i &=& \frac{1}{2} \Bigl(|\downarrow_{1,i} \uparrow_{2,i} \uparrow_{3,i} \uparrow_{4,i}\rangle - |\uparrow_{1,i} \downarrow_{2,i} \uparrow_{3,i} \uparrow_{4,i}\rangle \nonumber \\
	&+& |\uparrow_{1,i} \uparrow_{2,i} \downarrow_{3,i} \uparrow_{4,i}\rangle - |\uparrow_{1,i} \uparrow_{2,i} \uparrow_{3,i} \downarrow_{4,i}\rangle \Bigr), 
\label{triplet}
\end{eqnarray}
The strictly localized nature of the bound one-magnon eigenstate (\ref{triplet}) allows a straightforward derivation of exact many-magnon eigenstates with the character of Wigner magnon crystals. In particular, a regular alternation of the localized triplet state (\ref{triplet}) and fully polarized square plaquettes gives rise to a two-fold degenerate triplet-quintuplet ground state: 
\begin{eqnarray}\label{Eq:TQ(1-2)}
	|\text{TQ(1-2)}\rangle = 
	\left\{
	\begin{array}{l}
		\displaystyle \prod_{i=1}^{N_u/2} |\text{t}\rangle_{2i-1} \!\otimes\! 
		|\uparrow_{1,2i} \uparrow_{2,2i} \uparrow_{3,2i} \uparrow_{4,2i}\rangle \\[6pt]
		\displaystyle \prod_{i=1}^{N_u/2} 
		|\uparrow_{1,2i-1} \uparrow_{2,2i-1} \uparrow_{3,2i-1} \uparrow_{4,2i-1}\rangle 
		\!\otimes\!		|\text{t}\rangle_{2i}
	\end{array}\!\!\!\!\!\!\!\!\!.
	\right.
\end{eqnarray}
In the highly frustrated regime $J_2/J_1>2$, this exact eigenstate TQ(1-2) becomes the ground state just below the saturation field $h/J_1 = 2 J_2/J_1 + 4$. Interestingly, another Wigner magnon crystal formed entirely by the bound triplet states (\ref{triplet}) also constitutes an exact ground state of the fully frustrated spin-1/2 Heisenberg four-leg tube: 
\begin{eqnarray}\label{Eq:T(1)}
|\text{T(1)}\rangle =	\displaystyle \prod_{i=1}^{N_u} |\text{t}\rangle_{i}.  
\end{eqnarray}
The Wigner magnon crystal T(1) appears in the highly frustrated regime $J_2/J_1>2$ below the transition field $h/J_1 = 2 J_2/J_1 + 2$. The Wigner magnon crystals $\text{T(1)}$ and $\text{TQ(1-2)}$ emerging in the highly frustrated regime should manifest themselves macroscopically in the zero-temperature magnetization curves as intermediate plateaus at one-half and three-quarters of the saturation magnetization, respectively.  

\subsection{Bound two-magnon state and magnon crystals}
Although an analogous exact calculation of the complete spectrum in the two-magnon sector with $S_{tot}^z=2N_u-2$ is considerably more involved, one can readily verify that a plaquette-singlet state localized on an individual square plaquette constitutes an exact two-magnon eigenstate of the fully frustrated spin-1/2 Heisenberg four-leg tube. The corresponding plaquette-singlet state localized on the $i$th square plaquette is given by the eigenvector:
\begin{equation}\label{singlet}
	\begin{split}
		|s\rangle_i = \Biggl[ & \frac{1}{\sqrt{3}} \left( |\uparrow_{1,i} \downarrow_{2,i} \uparrow_{3,i} \downarrow_{4,i} \rangle + |\downarrow_{1,i} \uparrow_{2,i} \downarrow_{3,i} \uparrow_{4,i} \rangle \right) \\
		& - \frac{1}{\sqrt{12}} \Bigl( |\uparrow_{1,i} \uparrow_{2,i} \downarrow_{3,i} \downarrow_{4,i} \rangle + |\uparrow_{1,i} \downarrow_{2,i} \downarrow_{3,i} \uparrow_{4,i} \rangle \\
		& + |\downarrow_{1,i} \uparrow_{2,i} \uparrow_{3,i} \downarrow_{4,i} \rangle + |\downarrow_{1,i} \downarrow_{2,i} \uparrow_{3,i} \uparrow_{4,i} \rangle \Bigr) \Biggr].
	\end{split}
\end{equation}
The plaquette-singlet state (\ref{singlet}) is strictly localized on a single square plaquette and since its total spin is zero $T_i=0$, this bound two-magnon eigenstate completely decouples spins of a given square plaquette from all other spins. Consequently, other exact many-magnon eigenstates of the fully frustrated spin-1/2 Heisenberg four-leg tube can be constructed by placing the plaquette-singlet state (\ref{singlet}) on selected square plaquettes and such exact eigenstates provide another class of Wigner magnon crystals. In particular, the bound magnon crystal with the plaquette-singlet state (\ref{singlet}) on every unit cell gives rise to the fully fragmented singlet ground state:
\begin{eqnarray}\label{Eq:S(0)}
	|\text{S(0)}\rangle = \displaystyle \prod_{i=1}^{N_u} |\text{s}\rangle_{i}.
\end{eqnarray}
In the highly frustrated regime $J_2/J_1>2$, this exact eigenstate becomes the ground state for magnetic fields smaller than $h/J_1 = J_2/J_1$. Above this transition field, the other Wigner magnon crystal with character of a doubly-degenerate singlet-triplet state emerges due to a regular alternation of the plaquette-singlet state (\ref{singlet}) and the localized triplet state (\ref{triplet}):
\begin{eqnarray}\label{Eq:ST(0-1)}
|\text{ST(0-1)}\rangle &=& 
\left\{
\begin{array}{l}
	\displaystyle \prod_{i=1}^{N_u/2} |\text{s}\rangle_{2i-1} \,\otimes\, 
	|\text{t}\rangle_{2i} \\[6pt]
	\displaystyle \prod_{i=1}^{N_u/2} 
	|\text{t}\rangle_{2i-1} \,\otimes\, 
	|\text{s}\rangle_{2i}
\end{array}.
\right.
\end{eqnarray}
A subtle interplay between the magnetic field and spin frustration may give rise to another Wigner magnon crystal with character of a doubly-degenerate singlet-quintuplet state, which results from a
regular alternation of the plaquette-singlet state (\ref{singlet}) and fully polarized square plaquettes:
\begin{eqnarray}\label{Eq:SQ(0-2)}
	|\text{SQ(0-2)}\rangle = 
	\left\{
	\begin{array}{l}
		\displaystyle \prod_{i=1}^{N_u/2} |\text{s}\rangle_{2i-1} \,\otimes\, 
		|\uparrow_{1,2i} \uparrow_{2,2i} \uparrow_{3,2i} \uparrow_{4,2i}\rangle \\[6pt]
		\displaystyle \prod_{i=1}^{N_u/2} 
		|\uparrow_{1,2i-1} \uparrow_{2,2i-1} \uparrow_{3,2i-1} \uparrow_{4,2i-1}\rangle \,\otimes\, 
		|\text{s}\rangle_{2i}
	\end{array}\!\!\!\!\!\!\!\!\!\!\!\!.
	\right.
\end{eqnarray}
To conclude, the Wigner magnon crystals $\text{S(0)}$, $\text{ST(0-1)}$, and $\text{SQ(0-2)}$ should manifest themselves macroscopically in the zero-temperature magnetization curves as intermediate plateaus at zero, one-quarter, and one-half of the saturation magnetization, respectively.   

\subsection{Effective interacting lattice-gas model}
\label{sec:lattice-gass_model}

\begin{figure*}
	\centering
	\includegraphics[width=0.8\textwidth]{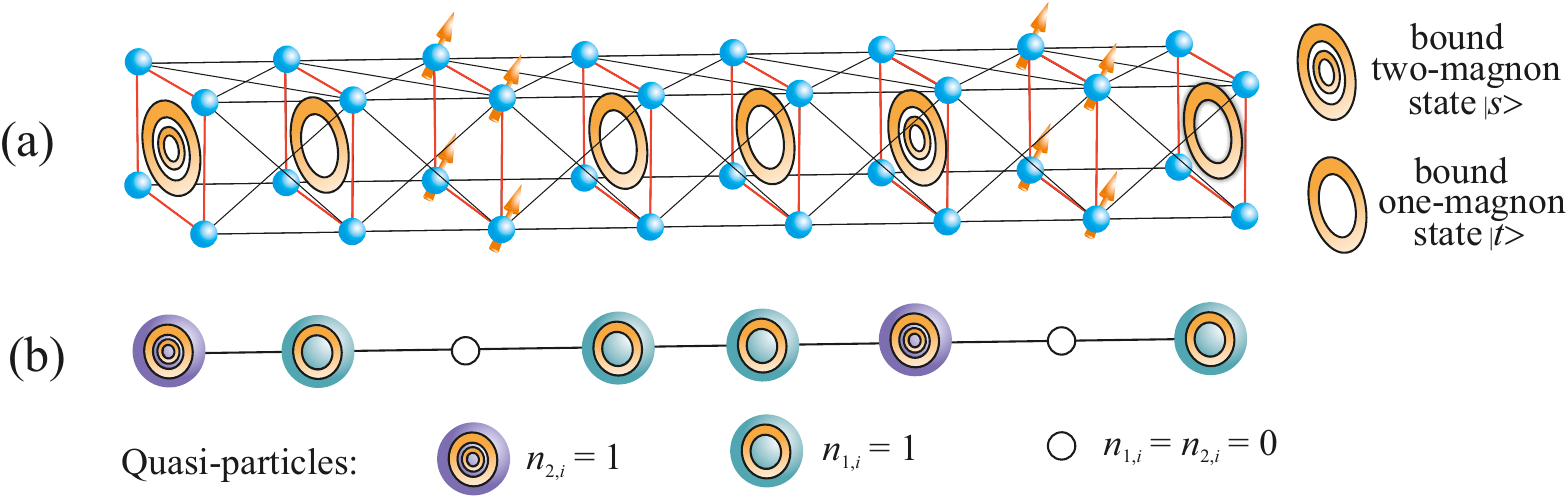}
		\vspace{-0.4cm}
	\caption{Schematic representation of a selected eigenstate of the fully frustrated spin-1/2 Heisenberg four-leg tube (a) and its equivalent representation within the effective two-component lattice-gas model (b). The single and double orange circles denote the bound one- and two-magnon states of a square plaquette, respectively, which are represented in the effective description by the two kinds of hard-core monomer quasi-particles illustrated by green and violet balls. Fully polarized square plaquettes correspond to empty (unoccupied) lattice sites.}
\label{fig:LGM_4LegL}
\end{figure*}

All five Wigner magnon crystals (\ref{Eq:TQ(1-2)}), (\ref{Eq:T(1)}), (\ref{Eq:S(0)}), (\ref{Eq:ST(0-1)}), and (\ref{Eq:SQ(0-2)}) become relevant in the moderately and highly frustrated parameter regime with a sufficiently high value of the interaction ratio $J_2/J_1$. To describe analytically the magnetic and thermodynamic behavior in the frustrated regime, we therefore map the original fully frustrated spin-1/2 Heisenberg four-leg tube onto an effective classical lattice-gas model involving two species of hard-core monomers, which represent the bound one- and two-magnon states (\ref{triplet}) and (\ref{singlet}) localized on individual square plaquettes. The corresponding effective Hamiltonian then reads:
\begin{eqnarray}\label{Eq:H_eff}
	\mathcal{H}_\mathrm{eff} & = & E_\mathrm{FM} - \mu_1 \sum_{i=1}^{N_u} n_{1,i} - \mu_2 \sum_{i=1}^{N_u} n_{2,i} \nonumber \\ 
	& & + \; J_1 \sum_{i=1}^{N_u} n_{1,i} n_{1,i+1}. 
\end{eqnarray}
Here, $E_\mathrm{FM} = N_u(4J_1 + 2J_2 - 2h)$ is the energy of the fully polarized ferromagnetic state $|\text{FM}\rangle$, while the occupation numbers $n_{1,i}, n_{2,i} \in \{0,1\}$ specify the absence or presence of the two quasi-particle species on the $i$th square plaquette. The corresponding chemical potentials $\mu_1$ and $\mu_2$ of these quasi-particles determine the energy gain associated with creating the bound one- and two-magnon states (\ref{triplet}) and (\ref{singlet}) on the fully polarized $|\text{FM}\rangle$ background:
\begin{itemize}
	\item[($i$)] $\mu_1 = 4J_1 + 2J_2 - h$ for the one-magnon state (\ref{triplet}).
	\item[($ii$)] $\mu_2 = 4J_1 + 3J_2 - 2h$ for the two-magnon state (\ref{singlet}).
\end{itemize}
The last term in the effective Hamiltonian (\ref{Eq:H_eff}) accounts for an energy cost $J_1$ associated with the repulsive interaction between the bound one-magnon states (\ref{triplet}) on two adjacent square plaquettes. Fig. \ref{fig:LGM_4LegL} schematically illustrates a representative eigenstate of the fully frustrated spin-1/2 Heisenberg four-leg tube with its equivalent representation in terms of the effective lattice-gas model.
The grand-canonical partition function of the effective lattice-gas model can then be written as:
\begin{eqnarray}\label{Eq:Z_eff}
	\mathcal{Z}_\mathrm{eff} \!&=&\! \exp(-\beta E_\mathrm{FM}) 
	\sum_{\{n_{1,i}\}} \sum_{\{n_{2,i}\}} \prod_{i=1}^{N_u} (1-n_{1,i} n_{2,i}) \nonumber \\
	\!&\times&\! \exp[\beta (\mu_1 n_{1,i} + \mu_2 n_{2,i} - J_1 n_{1,i} n_{1,i+1})].
\end{eqnarray}
The factor $(1-n_{1,i} n_{2,i})$ enforces the hard-core constraint, which excludes simultaneous occupation of the same square plaquette by the two quasiparticle species. Performing the summation over the occupation numbers $\{n_{2,i}\}$ associated with the bound two-magnon states reduces the partition function to that of an effective one-dimensional lattice-gas problem of interacting quasiparticles ascribed to the bound one-magnon states, whose distribution is given by the set of occupation numbers $\{n_{1,i}\}$ of the other quasiparticle species. The resulting partition function can be evaluated exactly using the transfer-matrix method:
\begin{eqnarray}\label{Eq:Z_eff_transfer}
	\mathcal{Z}_\mathrm{eff} & = & \exp(-\beta E_\mathrm{FM}) \sum_{\{n_{1,i}\}} 
	\prod_{i=1}^{N_u} T_{n_{1,i}, n_{1,i+1}},
\end{eqnarray}
where the transfer matrix $T_{n_{1,i}, n_{1,i+1}}$ is defined as:
\begin{eqnarray}\label{Eq:Tmatrix}
	T_{n_{1,i}, n_{1,i+1}} & = & \begin{pmatrix} 
		1 + {\rm e}^{\beta \mu_2} & 1 + {\rm e}^{\beta \mu_2} \\ 
		{\rm e}^{\beta \mu_1} & {\rm e}^{\beta (\mu_1 - J_1)} 
	\end{pmatrix}.
\end{eqnarray}
A straightforward diagonalization of the transfer matrix (\ref{Eq:Tmatrix}) yields the two eigenvalues: 
\begin{eqnarray}\label{Eq:eigenvalues}
	\lambda_{\pm} & = & \frac{1}{2} \Bigl[ (1 + {\rm e}^{\beta \mu_2}) + {\rm e}^{\beta (\mu_1 - J_1)} \\
	& & \pm \sqrt{ \left[ (1 + {\rm e}^{\beta \mu_2}) - {\rm e}^{\beta (\mu_1 - J_1)} \right]^2 + 4(1 + {\rm e}^{\beta \mu_2}) {\rm e}^{\beta \mu_1} } \, \Bigr], \nonumber 
\end{eqnarray}
in terms of which the partition function takes the form:
\begin{eqnarray}\label{Eq:Z_final}
	\mathcal{Z}_\mathrm{eff} & = & {\rm e}^{-\beta E_\mathrm{FM}} \bigl(\lambda_{+}^{N_u}+\lambda_{-}^{N_u}\bigr).
\end{eqnarray}
In the thermodynamic limit $N_u \to\infty$, the partition function is governed by the largest transfer-matrix eigenvalue $\lambda_+$ and the Gibbs free energy becomes:
\begin{eqnarray}\label{Eq:G_eff}
	\mathcal{G}_\mathrm{eff} & = & E_\mathrm{FM} - N_u k_\mathrm{B}T \ln \lambda_{+}.
\end{eqnarray}
The magnetization, magnetic susceptibility, entropy, and specific heat can then be obtained from the standard thermodynamic relations:
\begin{eqnarray}\label{Eq:thermo_relations}
	&& m_\mathrm{eff} = -\left(\frac{\partial \mathcal{G}_\mathrm{eff}}{\partial h}\right)_{T},\quad \chi_\mathrm{eff} =  \left(\frac{\partial m_\mathrm{eff}}{\partial h}\right)_{T},\nonumber \\
	&& S_\mathrm{eff} =  -\left(\frac{\partial \mathcal{G}_\mathrm{eff}}{\partial T}\right)_{h},\quad
	C_\mathrm{eff} =  T\left(\frac{\partial S_\mathrm{eff}}{\partial T}\right)_{h}.
\end{eqnarray}

\subsection{Density matrix renormalization group method}

The Hamiltonian (\ref{Eq:4ll_ham}) of the fully frustrated spin-1/2 Heisenberg four-leg tube commutes with the square of the local composite spin operator of each square plaquette $\hat{\mathbf{T}}_{i} = \sum_{j=1}^{4} \hat{\mathbf{S}}_{i,j}$, i.e. $[\hat{H}, \hat{\mathbf{T}}_{i}^2] = 0$. Consequently, the total spin of each square plaquette $T_i$ represents a locally conserved quantity with a well-defined quantum spin numbers $T_i \in \{0, 1, 2\}$. This local conservation law allows for a substantial reduction in the computational complexity of the original problem. In terms of the composite spin operators, the Hamiltonian (\ref{Eq:4ll_ham}) of the fully frustrated spin-1/2 Heisenberg four-leg tube can be rewritten in the following equivalent form:
\begin{eqnarray}
	\hat{H} = \sum_{i=1}^{N_u} \left[ J_1 \hat{\mathbf{T}}_i \cdot \hat{\mathbf{T}}_{i+1} + \frac{J_2}{2} 
	\left( \hat{\mathbf{T}}_{i}^2 - \hat{\mathbf{S}}_{13,i}^2 - \hat{\mathbf{S}}_{24,i}^2 \right) - h \hat{T}_i^z \right], \nonumber \\
	\label{hameff}
\end{eqnarray}
where $\hat{\mathbf{S}}_{13,i} = \hat{\mathbf{S}}_{i,1} + \hat{\mathbf{S}}_{i,3}$ and $\hat{\mathbf{S}}_{24,i} = \hat{\mathbf{S}}_{i,2} + \hat{\mathbf{S}}_{i,4}$ denote the composite spin operators for two spin pairs from opposite corners of the $i$th square plaquette. The new representation of the Hamiltonian (\ref{hameff}) of the fully frustrated spin-1/2 Heisenberg four-leg tube can be interpreted as the antiferromagnetic Heisenberg chain with the coupling constant $J_1$ in a magnetic field $h$, whose local spin quantum numbers may vary from site to site and take one of the three possible values 
$T_i \in \{0, 1, 2\}$. In addition, the term incorporating the coupling constant $J_2$ provides a trivial shift of energy, which depends on the overall distribution of quantum spin numbers along the whole chain. 

With regard to a translational invariance and the antiferromagnetic nearest-neighbor coupling, all ground states of the fully frustrated spin-1/2 Heisenberg four-leg tube can be derived from the Hamiltonian (\ref{hameff}) by restricting possible combinations of spin magnitudes only to the antiferromagnetic Heisenberg chains with either uniform or regularly alternating sequences of the composite spins $T_i$. Recall that the plaquette-singlet state (\ref{singlet}) as the lowest-energy eigenstate of a square plaquette for the specific value of the composite quantum spin number $T_i=0$ breaks all inter-plaquette spin-spin correlations, which in turn leads to a fully fragmented singlet S(0) ground state given by Eq. (\ref{Eq:S(0)}) when considering $T_i=0$ for all square plaquettes. Similarly, one recovers from the Hamiltonian (\ref{hameff}) the two-fold degenerate singlet-triplet ST(0-1) and singlet-quintuplet SQ(0-2) ground states given by Eqs. (\ref{Eq:ST(0-1)}) and (\ref{Eq:SQ(0-2)}) when considering the regular alternation of $T_i=0$ with either $T_i=1$ or $T_i=2$, respectively. 

The remaining three relevant cases correspond to the Heisenberg spin chain (\ref{hameff}) with uniform composite spins $T_i=2$, uniform composite spins $T_i=1$, and regularly alternating composite spins $T_i=2$ and $T_i=1$. To determine their lowest-energy eigenstates, we performed the density matrix renormalization group (DMRG) calculations implemented within the Algorithms and Libraries for Physics Simulations (ALPS) project \cite{ALPS2011} for systems of up to $N_u=120$ unit cells corresponding to the total number of $N=480$ spins. The sufficiently large system sizes considered substantially reduce finite-size effects and provide a reliable approximation to the thermodynamic limit. The DMRG calculations for the antiferromagnetic Heisenberg spin chain (\ref{hameff}) with uniform composite spins $T_i=2$ reveal two additional ground states in addition to the fully polarized ferromagnetic phase. The first ground state is the Haldane phase H(2) with a relatively small energy gap, while the gapless Tomonaga-Luttinger quantum spin liquid TL(2) ground state evolves upon closing the Haldane gap. The DMRG calculations for the antiferromagnetic Heisenberg spin chain (\ref{hameff}) with uniform composite spins $T_i=1$ supports the other Haldane phase H(1), the Tomonaga-Luttinger liquid TL(1), and the fully polarized state T(1). However, comparison with the lowest energies of the competing eigenstates shows that neither H(1) nor TL(1) becomes the ground state of the original fully frustrated spin-1/2 Heisenberg four-leg tube. Hence, only the fully polarized state of the antiferromagnetic spin-1 Heisenberg chain T(1) represents the actual ground state, which can be identified with the Wigner magnon crystal resulting from a full condensation of the bound triplet states (\ref{Eq:T(1)}). The last two ground states of the fully frustrated spin-1/2 Heisenberg four-leg tube were derived from the DMRG calculations of the ferrimagnetic Heisenberg spin chain (\ref{hameff}) with the regularly alternating composite spins $T_i=2$ and $T_i=1$. The first ground state is the gapless Tomonaga-Luttinger quantum spin liquid TL(2-1) phase, whereas the second one corresponds to the fully polarized state of this ferrimagnetic mixed-spin chain. The latter corresponds to the Wigner magnon crystal (\ref{Eq:TQ(1-2)}) with regularly alternating triplet and quintuplet states on the square plaquettes.  

\subsection{Exact diagonalization method}
To validate the analytical predictions derived from the effective interacting lattice-gas model and the numerical DMRG results, we performed exact diagonalization of the original Hamiltonian (\ref{Eq:4ll_ham}) of the fully frustrated spin-1/2 Heisenberg four-leg tube using routines from the ALPS project \cite{ALPS2011}. To this end, we employed the routine '\textit{fulldiag}' for complete exact diagonalization (ED) of the fully frustrated spin-1/2 Heisenberg four-leg tube with up to $N_u=4$ unit cells ($N=16$ spins) and the routine '\textit{sparsediag}' for exact diagonalization based on the Lanczos algorithm (EDL) for systems with up to $N_u=6$ unit cells ($N=24$ spins). Although these exact results are restricted to relatively small system sizes, they still provide a useful independent benchmark for both the approximate analytical results derived from the effective interacting lattice-gas model and the DMRG data obtained only from a subset of all possible quantum spin chains involved in the Hamiltonian (\ref{hameff}).

\section{Ground-state phase diagram and magnetization curves}\label{sec:GSPD}

\begin{figure}[t]
	\centering
	\includegraphics[width=0.5\textwidth]{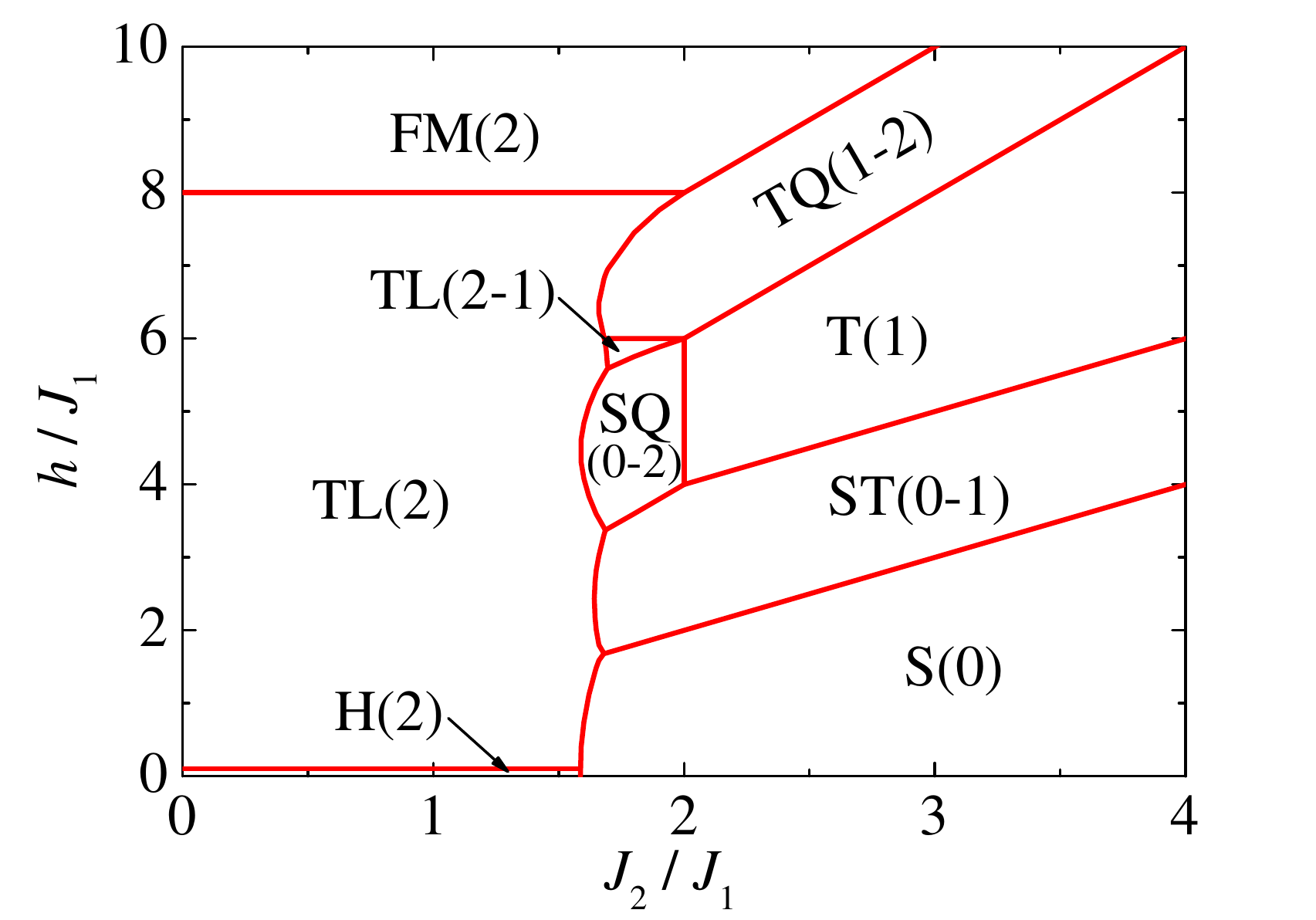} 
	\vspace{-0.3cm}
	\caption{Ground-state phase diagram of the fully frustrated spin-1/2 Heisenberg four-leg tube in the $J_2/J_1-h/J_1$ plane obtained from DMRG calculations for a finite-size system with 
	$N_u=120$ unit cells ($N=480$ spins). The individual ground states are denoted as follows: S(0) - singlet phase, ST(0-1) - singlet-triplet phase, T(1) - triplet phase, TQ(1-2) - triplet-quintuplet phase, SQ(0-2) - singlet-quintuplet phase, FM(2) - ferromagnetic phase, H(2) - Haldane phase, TL(2) and TL(2-1) - Tomonaga-Luttinger quantum spin liquids. The numbers in round brackets specify the composite quantum spin numbers $T_i$ of the square plaquettes.}
	\label{fig:GSPD}
\end{figure}

We begin by presenting the ground-state phase diagram of the fully frustrated spin-$1/2$ Heisenberg four-leg tube, which is displayed in Fig.~\ref{fig:GSPD} in the plane of the interaction ratio $J_2/J_1$ versus the magnetic field $h/J_1$. The phase diagram was established using DMRG simulations for a finite-size system with $N_u=120$ unit cells corresponding to $N=480$ spins. The ground-state phase diagram reveals a rich variety of quantum phases that can be broadly classified into three regimes: the unfrustrated to weakly frustrated regime $J_2/J_1 \ll 2$, the moderately frustrated regime $J_2/J_1 \lesssim 2$, and the highly frustrated regime $J_2/J_1 \geq 2$.

In the unfrustrated to weakly frustrated regime $J_2/J_1 \ll 2$, the system exhibits three ground states inherently connected to the effective antiferromagnetic spin-2 Heisenberg chain (\ref{hameff}) with uniform composite spins $T_i = 2$. At low magnetic fields, the ground state is the gapped Haldane phase H(2) followed by the gapless Tomonaga-Luttinger quantum spin liquid TL(2) before reaching the fully polarized ferromagnetic phase FM(2) at the saturation field. In the moderately frustrated regime $J_2/J_1 \lesssim 2$, two additional phases emerge at intermediate magnetic fields. Another gapless Tomonaga-Luttinger quantum spin liquid TL(2-1) results from the ferrimagnetic Heisenberg spin chain (\ref{hameff}) with regularly alternating composite spins $T_i = 2$ and $1$. In addition, this regime also hosts the Wigner magnon crystal SQ(0-2) characterized by a regular alternation of plaquette singlets ($T_i = 0$) and quintuplets ($T_i = 2$). The highly frustrated regime $J_2/J_1 \geq 2$ is dominated by the four Wigner magnon crystals involving bound one-magnon and/or two-magnon states: the singlet phase S(0), the singlet-triplet phase ST(0-1), the triplet phase T(1), and the triplet-quintuplet phase TQ(1-2). 

\begin{figure*}[t]
	\centering
	\includegraphics[scale=0.22,trim=20 0 40 0, clip]{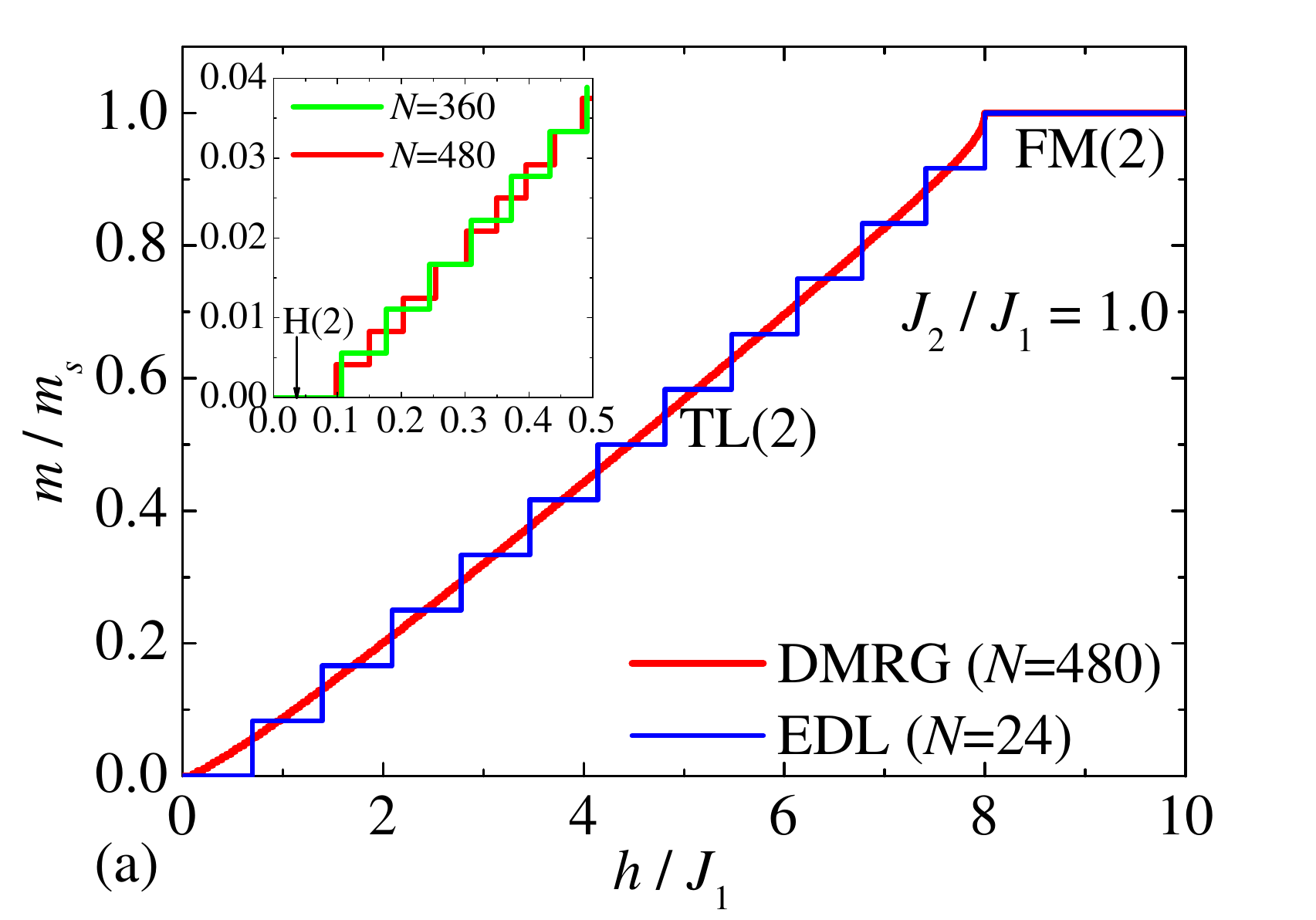}
	\includegraphics[scale=0.22,trim=20 0 40 0, clip]{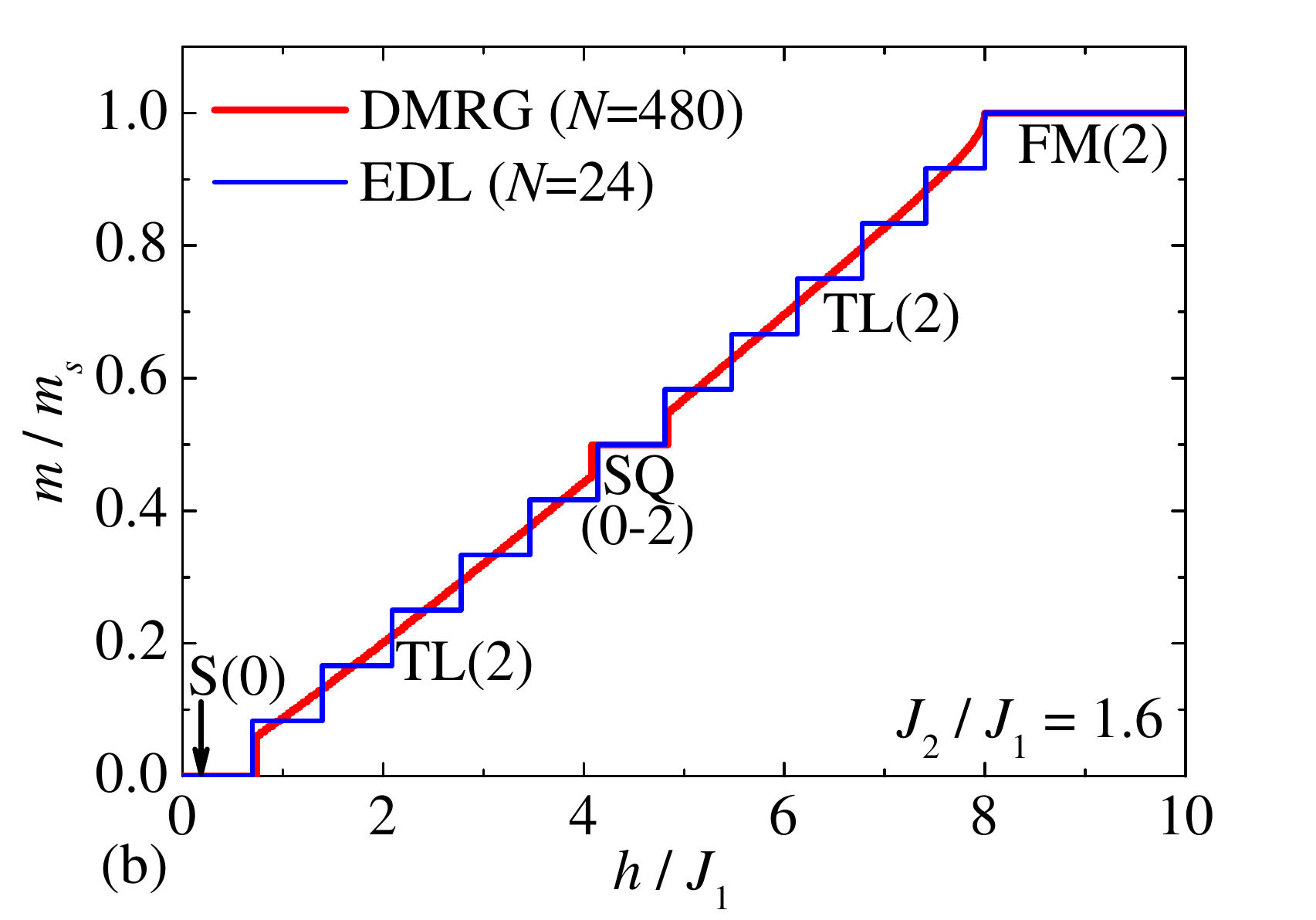}
	\includegraphics[scale=0.22,trim=20 0 40 0, clip]{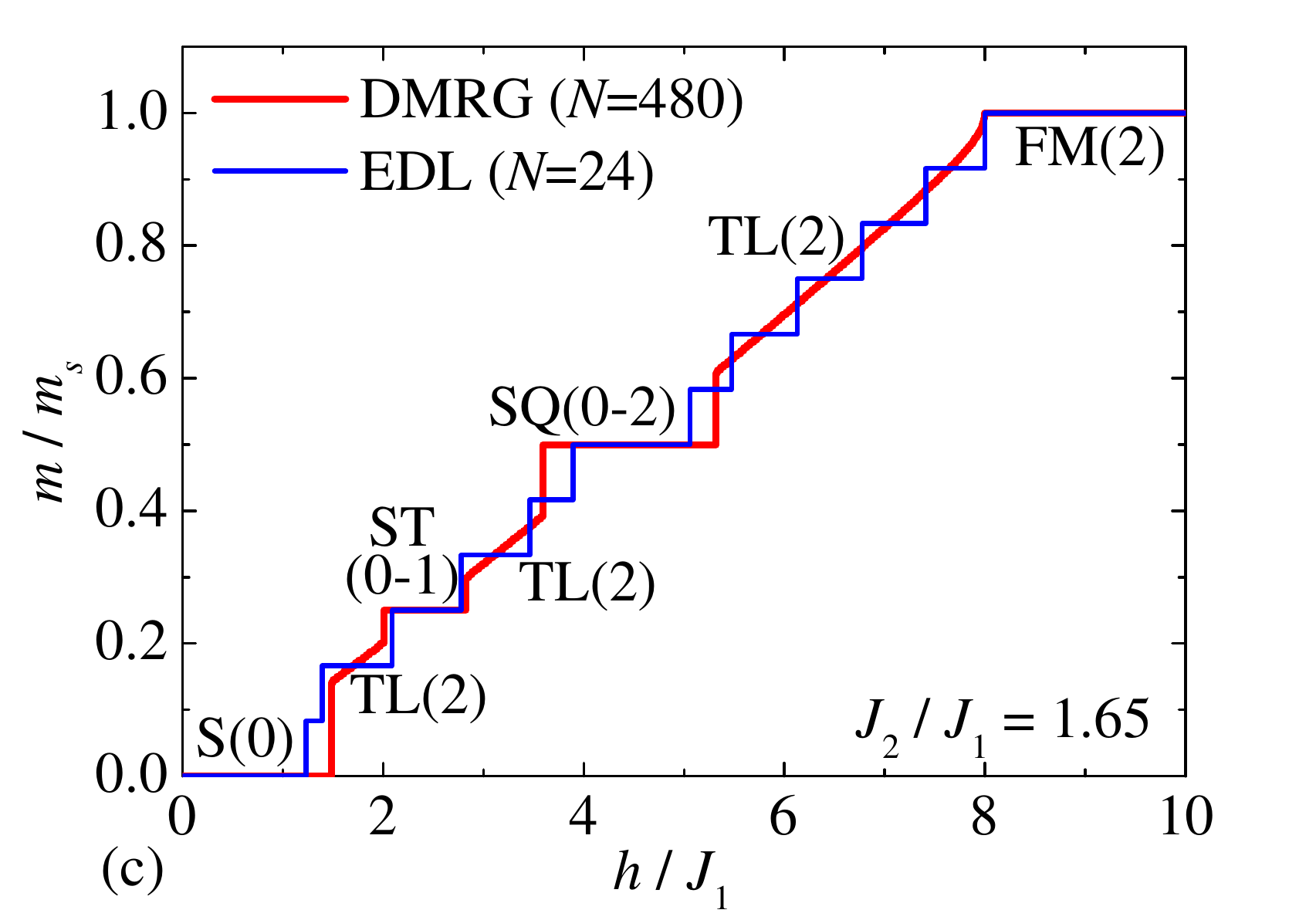}
	\includegraphics[scale=0.22,trim=20 25 60 0, clip]{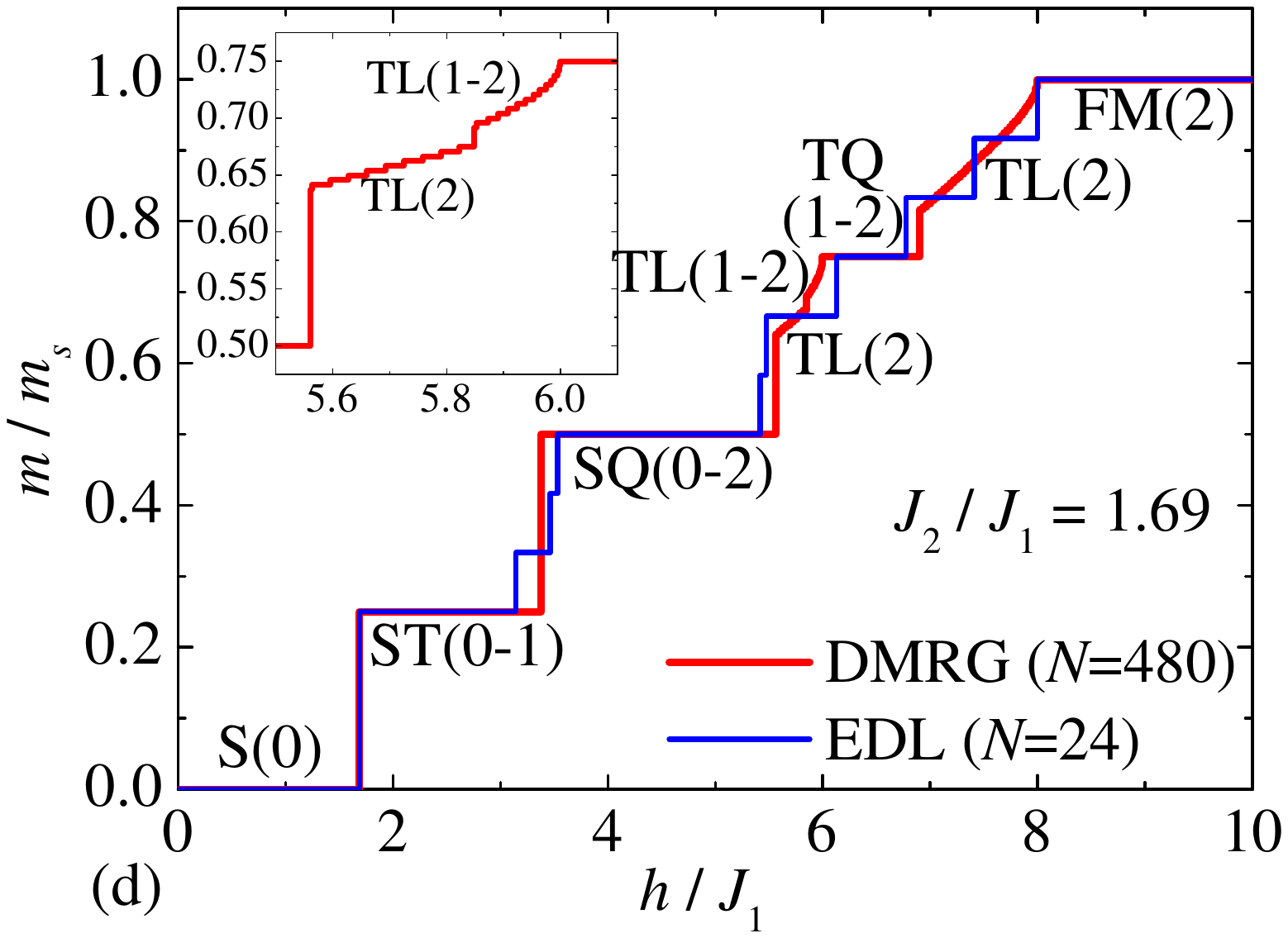}
	\includegraphics[scale=0.22,trim=20 0 40 0, clip]{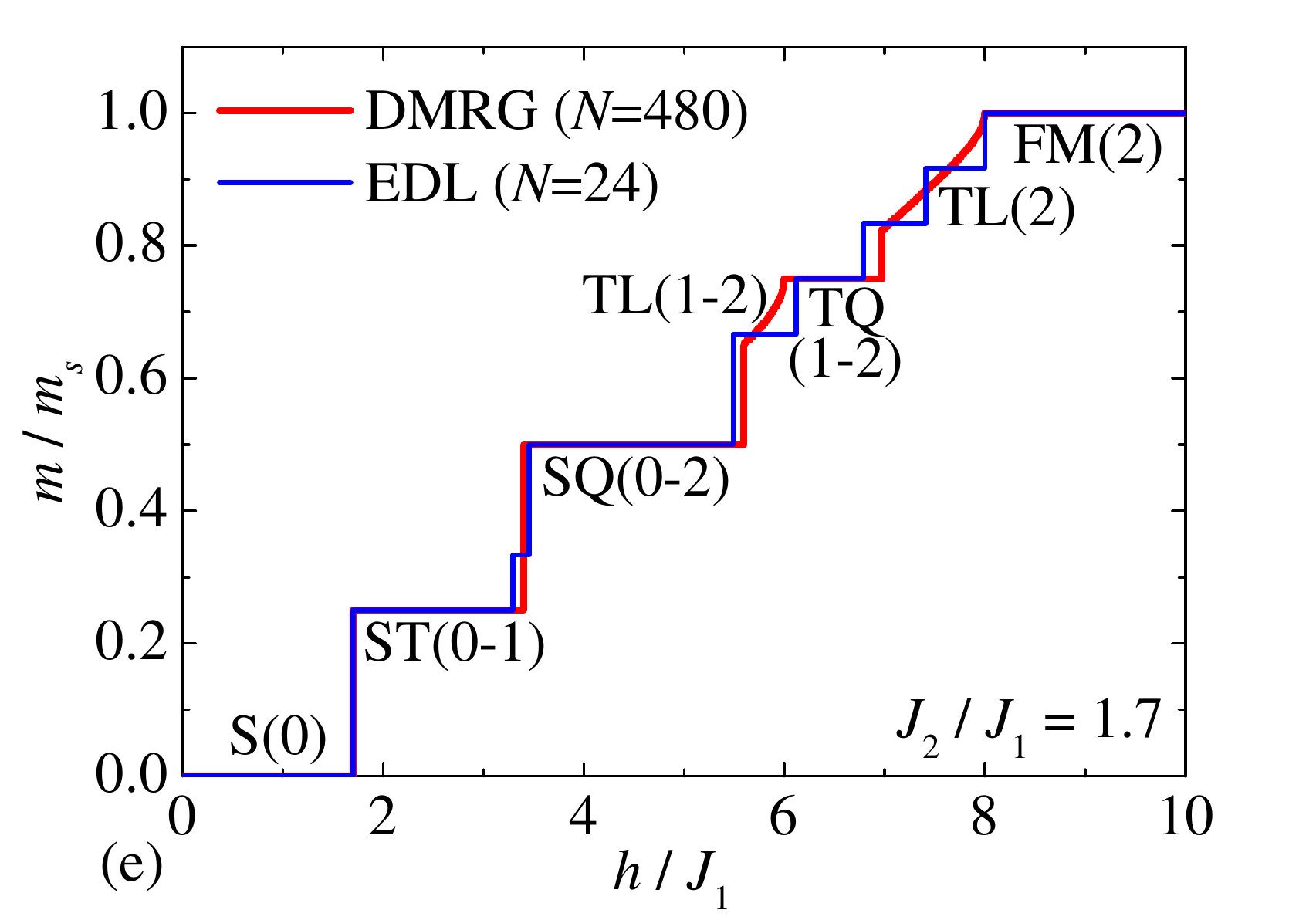}
	\includegraphics[scale=0.22,trim=20 0 40 0, clip]{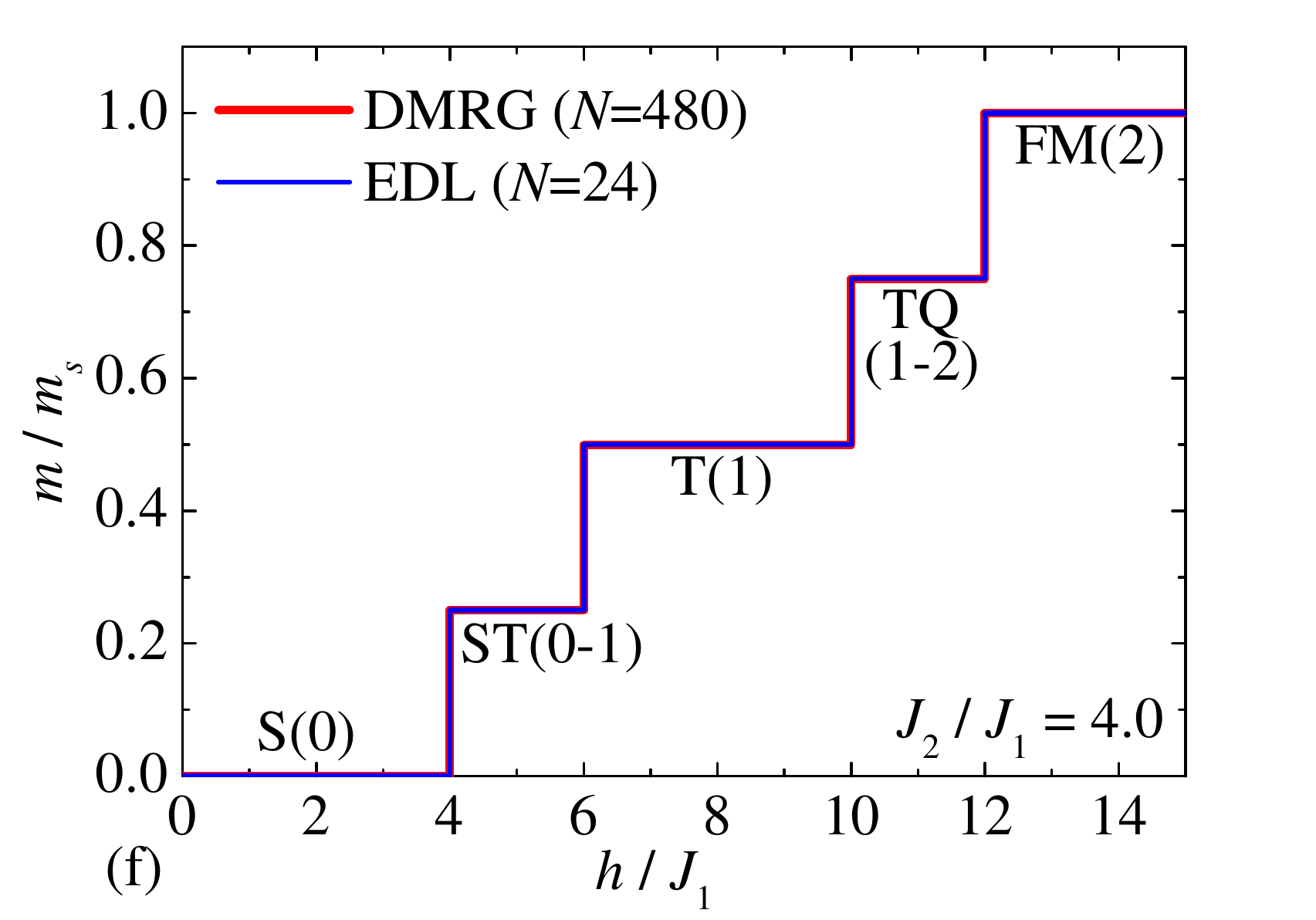}
	\caption{Zero-temperature magnetization curves of the fully frustrated spin-1/2 Heisenberg four-leg tube calculated by EDL (blue lines) and DMRG (red lines) methods for six selected values of the interaction ratio: (a) $J_{2}/J_{1}=1.0$, (b) $J_{2}/J_{1}=1.6$, (c) $J_{2}/J_{1}=1.65$, (d) $J_{2}/J_{1}=1.67$, (e) $J_{2}/J_{1}=1.7$, and (f) $J_{2}/J_{1}=4.0$. The magnetization is normalized with respect to its saturation value $m/m_{\mathrm{s}}$, the EDL and DMRG results are presented for $N=24$ and $N=480$ spins, respectively. The inset in panel (a) illustrates finite-size effects in the low-field DMRG magnetization curves for the systems involving $N=360$ and $N=480$ spins, while the inset in panel (d) provides an enlarged view of the discontinuous magnetization jump at the transition between the two gapless phases TL(2) and TL(1-2).}
	\label{fig:Mz_h_T0}
\end{figure*}

To further corroborate the ground-state phase diagram, Fig.~\ref{fig:Mz_h_T0} presents the zero-temperature magnetization curves for six representative values of the interaction ratio $J_2/J_1$. A comparison between the EDL results for $N=24$ spins and DMRG results for $N=480$ spins brings insight into finite-size effects. The sequence of magnetization plateaus and jumps directly reflects the phases identified in the ground-state phase diagram and the field-driven transitions between them emerging along their respective ground-state boundaries, respectively. 

For $J_2/J_1=1.0$ in the weakly frustrated regime, the magnetization curve exhibits a narrow zero-magnetization plateau associated with the gapped Haldane phase H(2) as depicted in Fig.~\ref{fig:Mz_h_T0}(a) and its inset. The zero magnetization plateau terminates at a field-driven quantum phase transition to the gapless Tomonaga-Luttinger quantum spin liquid TL(2), in which the magnetization increases continuously with the magnetic field in the thermodynamic limit. The finite-size analysis illustrated in the inset of Fig. \ref{fig:Mz_h_T0}(a) clearly reveals that the magnetization steps within the quantum spin-liquid regime gradually decrease with increasing the system size and ultimately vanish in the thermodynamic limit, whereas the width of the zero-magnetization plateau remains nearly unchanged and gradually converges to a finite value corresponding to the Haldane gap (the zero magnetization step is almost twice as large as the other magnetization steps for largest system size considered with $N=480$ spins). Above the critical field at which the Haldane gap closes, the magnetization increases continuously with the magnetic field until it reaches a full saturation at the second field-driven quantum phase transition toward the fully polarized ferromagnetic phase FM(2).

A qualitatively different behavior emerges in the moderately frustrated regime $J_2/J_1 \lesssim 2$. For the interaction ratio $J_2/J_1=1.6$, the zero-temperature magnetization curve exhibits a pronounced zero-magnetization plateau corresponding to the Wigner magnon crystal S(0), in which all square plaquettes are in the singlet state [Fig.~\ref{fig:Mz_h_T0}(b)]. This plateau extends over a substantially wider field range due to much higher energy gap of the magnon crystal S(0) compared to that of the Haldane phase H(2) observed for $J_2/J_1=1.0$.  Another distinction is that the system enters the gapless TL(2) phase at the discontinuous rather than the continuous field-induced phase transition when closing this energy gap. Interestingly, the other Wigner magnon crystal SQ(0-2) characterized by a regular alternation of plaquette singlets and fully polarized square plaquettes emerges within the field range otherwise occupied by the TL(2) phase and gives rise to an intermediate plateau at one-half of the saturation magnetization. Upon a further increase of the spin frustration to $J_2/J_1=1.65$ and $1.67$, additional magnetization plateaus develop as illustrated in Figs.~\ref{fig:Mz_h_T0}(c) and \ref{fig:Mz_h_T0}(d). In particular, the intermediate one-quarter and three-quarters plateaus correspond to the Wigner magnon crystals ST(0-1) and TQ(1-2), respectively. At the same time, the field range of the gapless TL(2) phase gradually shrinks, while the second gapless phase TL(2-1) emerges in a narrow field interval. Remarkably, the phase transition between the two gapless critical quantum spin liquids TL(2) and TL(2-1) is discontinuous as evidenced by the finite jump in the magnetization shown in the inset of Fig.~\ref{fig:Mz_h_T0}(d). The last example from the moderately frustrated regime shown in Fig.~\ref{fig:Mz_h_T0}(e) for the interaction ratio $J_2/J_1=1.7$ illustrates how the gapless quantum spin liquid TL(2-1) fully suppresses the TL(2) quantum spin liquid at intermediate magnetic fields, which are delimited by the two gapped SQ(0-2) and TQ(1-2) phases corresponding to the intermediate one-half and three-quarters magnetization plateaus, respectively.  

Finally, the zero-temperature magnetization curves have very distinct character in the highly frustrated regime $J_2/J_1 \geq 2.0$ as illustrated in Fig.~\ref{fig:Mz_h_T0}(f) for the particular case with $J_2/J_1 = 4.0$. In this regime, both gapless quantum spin liquids and continuous field-induced quantum phase transitions closely associated with their onset and breakdown disappear completely. The zero-temperature magnetization curves are instead characterized by four robust intermediate magnetization plateaus separated by discontinuous magnetization jumps. With increasing magnetic field, the fully frustrated spin-1/2 Heisenberg four-leg tube successively realizes the Wigner magnon crystals S(0), ST(0-1), T(1), and TQ(1-2) producing pronounced intermediate magnetization plateaus at zero, one-quarter, one-half, and three-quarters of the saturation magnetization before reaching the full saturation. In contrast to the moderately frustrated regime, the intermediate one-half plateau now corresponds to the Wigner magnon crystal T(1) composed entirely of bound one-magnon (triplet) states rather than a regular alternation of the singlet and fully polarized square plaquettes realized within the Wigner magnon crystal SQ(0-2).  

\section{Finite-temperature behavior in the highly frustrated regime}
\label{sec:magnetothermodynamics}

\begin{figure*}[t]
	\centering
	\includegraphics[scale=0.38,trim=0 0 0 0, clip]{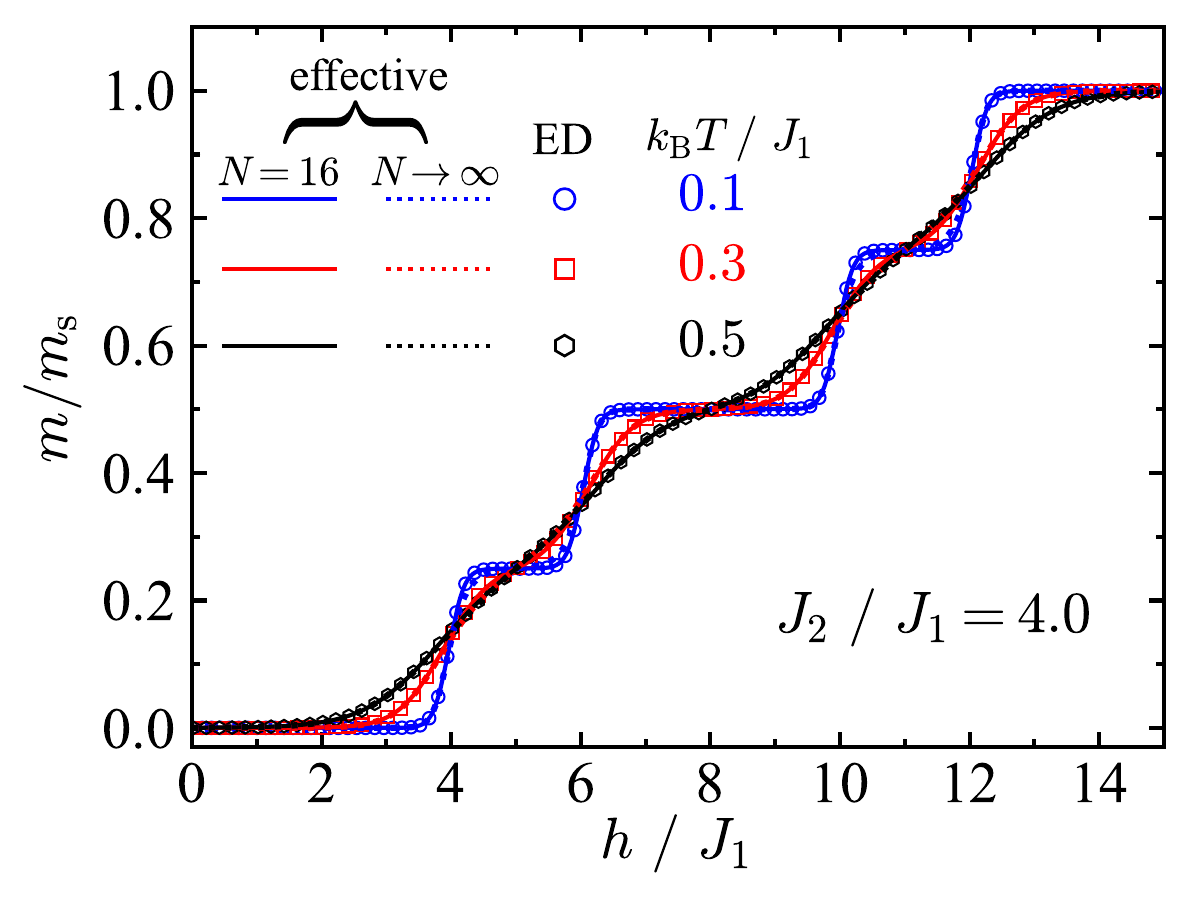}
	\includegraphics[scale=0.38,trim=0 0 0 0, clip]{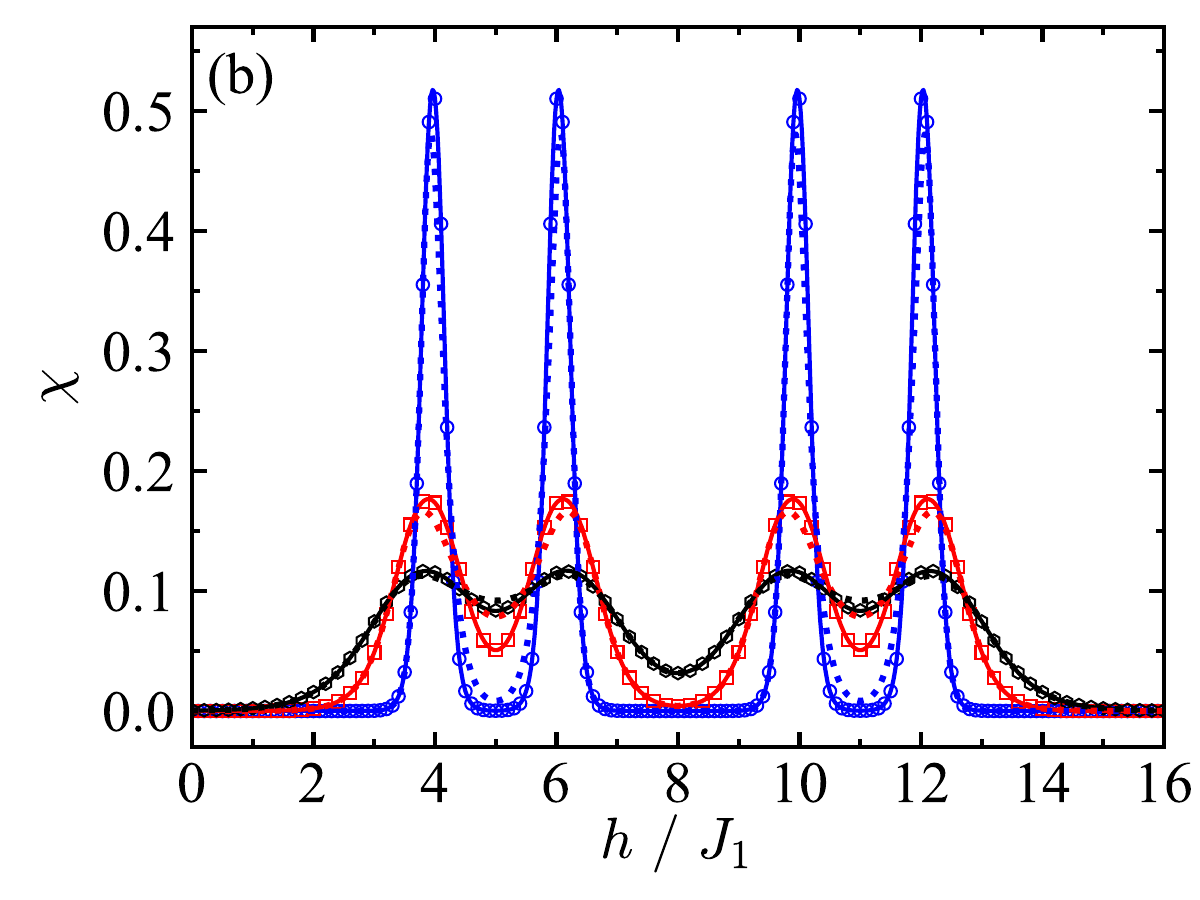}
	\includegraphics[scale=0.38,trim=0 0 0 0, clip]{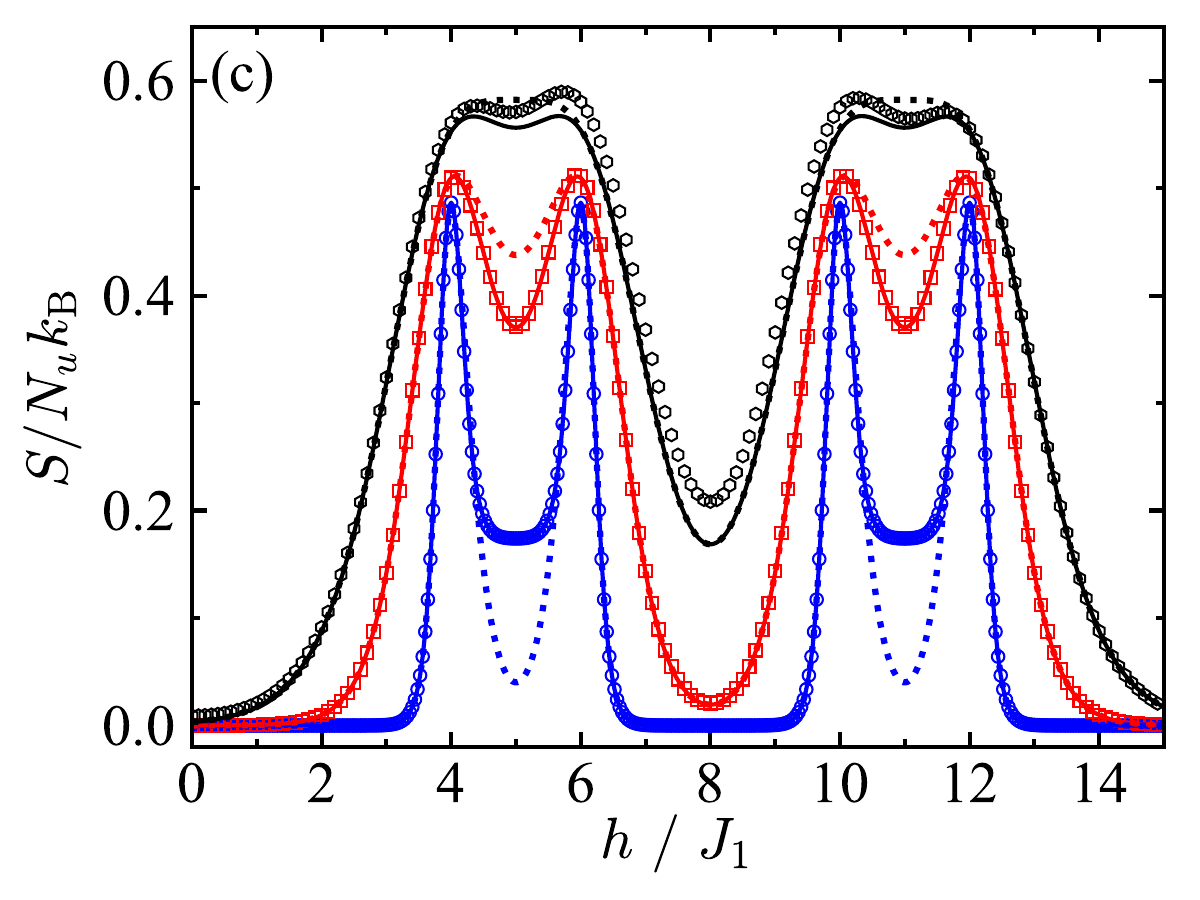}
	\includegraphics[scale=0.38,trim=0 0 0 0, clip]{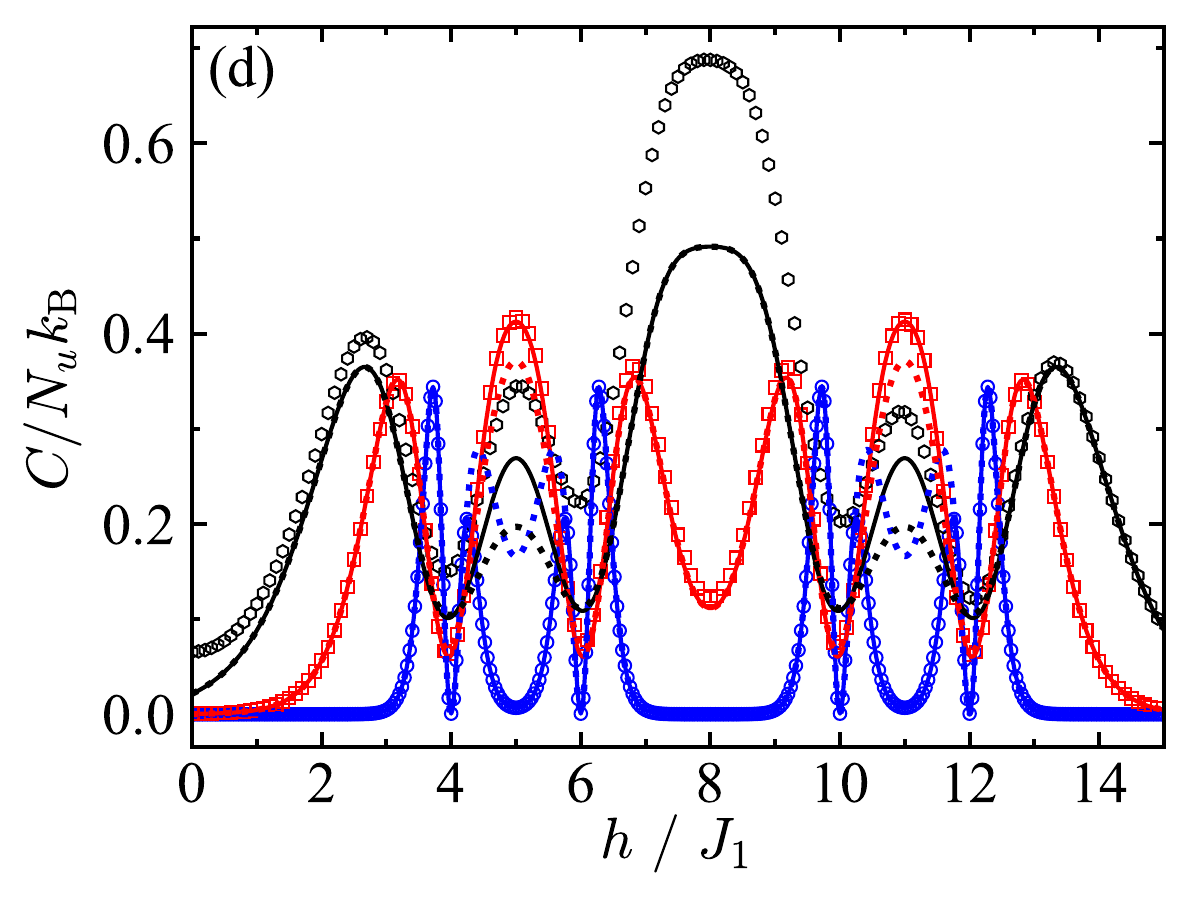}
	\caption{Finite-temperature properties of the fully frustrated spin-1/2 Heisenberg four-leg tube for one representative value of the interaction ratio $J_{2}/J_{1}=4$ and three different temperatures $k_\text{B}T/J_{1} = 0.1$, $0.3$, and $0.5$:
	(a) the magnetization normalized with respect to its saturation value $m/m_{\mathrm{s}}$; 
	(b) the magnetic susceptibility $\chi$; (c) the entropy per unit cell $S/N_uk_{\mathrm{B}}$; 
	(d) the specific heat per unit cell $C/N_u k_{\mathrm{B}}$. Symbols display the results of the full ED of the original model (\ref{Eq:4ll_ham}) for the number of unit cells $N_u=4$ (i.e. $N=16$ spins), while solid and dotted lines display the results derived from the effective interacting lattice-gas model (\ref{Eq:H_eff}) for the number of unit cells $N_u=4$ and $N_u \to \infty$, respectively.}
	\label{fig:Mz_S_ED_LM_J4}
\end{figure*}

The presence of the Wigner magnon crystals S(0), ST(0-1), T(1), and TQ(1-2) composed entirely of the bound one-magnon (\ref{triplet}) and/or two-magnon (\ref{singlet}) states allows a straightforward interpretation of finite-temperature properties of the fully frustrated spin-$1/2$ Heisenberg four-leg tube within the effective interacting lattice-gas model developed in Sec. \ref{sec:lattice-gass_model}. To assess the accuracy of this effective description, Figure~\ref{fig:Mz_S_ED_LM_J4} presents a comparison between the finite-temperature magnetic and thermodynamic quantities as obtained for the fully frustrated spin-$1/2$ Heisenberg four-leg tube from its full ED with those predicted by the effective lattice-gas description. The ED results correspond to a finite-size system with $N_u=4$ unit cells ($N=16$ spins), whereas the effective lattice-gas model was evaluated both for the same system size $N_u=4$ as well as in the thermodynamic limit $N_u\rightarrow\infty$. 

The isothermal magnetization curves shown in Fig.~\ref{fig:Mz_S_ED_LM_J4}(a) reveal excellent agreement between the full ED results and the predictions of the effective interacting lattice-gas model. Remarkably, the perfect agreement persists over the entire magnetic-field range and remains valid even at relatively elevated temperatures up to $k_{\mathrm B}T/J_{1}=0.5$. The perfect match between the numerical and analytical results demonstrates that the effective lattice-gas formulation captures the relevant low-energy excitations governing the magnetization process in the highly frustrated regime.  In particular, it correctly reproduces the nature and size of intermediate magnetization plateaus as well as gradual temperature-induced smearing of the discontinuous zero-temperature magnetization jumps between them. In addition, the close overlap between the finite-size and thermodynamic-limit solutions of the effective model also indicates negligible finite-size effects on the magnetization. The magnetic-field dependences of the magnetic susceptibility presented in Fig.~\ref{fig:Mz_S_ED_LM_J4}(b) provide complementary insights into these observations. The susceptibility maxima highlight rapid changes in the magnetization near all zero-temperature field-induced phase transitions, whereby the positions and overall shapes of these peaks are well reproduced by the effective lattice-gas model.  With increasing temperature, the susceptibility maxima become progressively lower and broader. The effective lattice-gas description quantitatively captures all these trends up to the relatively high temperature $k_{\mathrm B}T/J_{1}=0.5$. Although the difference between the finite-size ($N_u=4$) and thermodynamic-limit ($N_u \to \infty$) solutions still remains relatively small, finite-size effects are somewhat more pronounced than for the magnetization particularly at low temperatures but they generally decrease with increasing temperature. 

A somewhat different behavior is observed in the isothermal field dependences of the magnetic entropy and specific heat, which are depicted in Figs.~\ref{fig:Mz_S_ED_LM_J4}(c) and \ref{fig:Mz_S_ED_LM_J4}(d), respectively. The effective lattice-gas model accurately reproduces the ED results at low up to moderate temperatures $k_{\mathrm B}T/J_{1}=0.1$ and $0.3$, whereas noticeable deviations appear at higher temperatures such as $k_{\mathrm B}T/J_{1}=0.5$ particularly in the extrema of both these thermodynamic quantities. The observed discrepancies originate from higher-energy thermal excitations that are not included in the effective lattice-gas description, which is constructed from a restricted set of low-energy localized-magnon states. Owing to this fact, the quantitative accuracy of the interacting lattice-gas approach naturally decreases as temperature increases and additional excited states begin to contribute.  Nevertheless, the effective lattice-gas model still quantitatively captures the overall field dependence of the entropy and specific heat at low up to moderate temperatures and at least qualitatively at higher temperatures. The observed entropy maxima reflect accumulation of the low-energy states near the field-driven phase transitions, whereas the nonzero entropy plateau $S/N_u k_{\rm B} = \frac{1}{4} \ln 2 \approx 0.173$ observed for $N_u=4$ signals two-fold degeneracy of the two symmetry-broken Wigner magnon crystals ST(0-1) and TQ(1-2). Since the total degeneracy remains twofold independently of the system size, this latter nonzero value is only a finite-size artifact and the entropy (per unit cell) should vanish in the thermodynamic limit at low enough temperatures. This behavior is consistently reproduced by the thermodynamic-limit solution of the effective lattice-gas model, which illustrate how these features evolve with increasing the system size. Similar temperature trends and finite-size effects are also observed in the field dependences of the specific heat. The asymmetric double-peak structure of the specific heat reflects field-driven phase transition between the nondegenerate Wigner nmagnon crystal S(0) or T(1) and two-fold degenerate magnon crystal ST(0-1) or TQ(1-2). The most pronounced finite-size effects in the specific-heat can be detected at low temperatures, whereas the data $N_u=4$ and $N_u \to \infty$ data become nearly indistinguishable at higher temperatures.  

\section{Quantum Stirling heat engine}
\label{sec:heatengine}

\begin{figure}[t]
	\centering
	\includegraphics[scale=0.37,trim=0 0 0 0, clip]{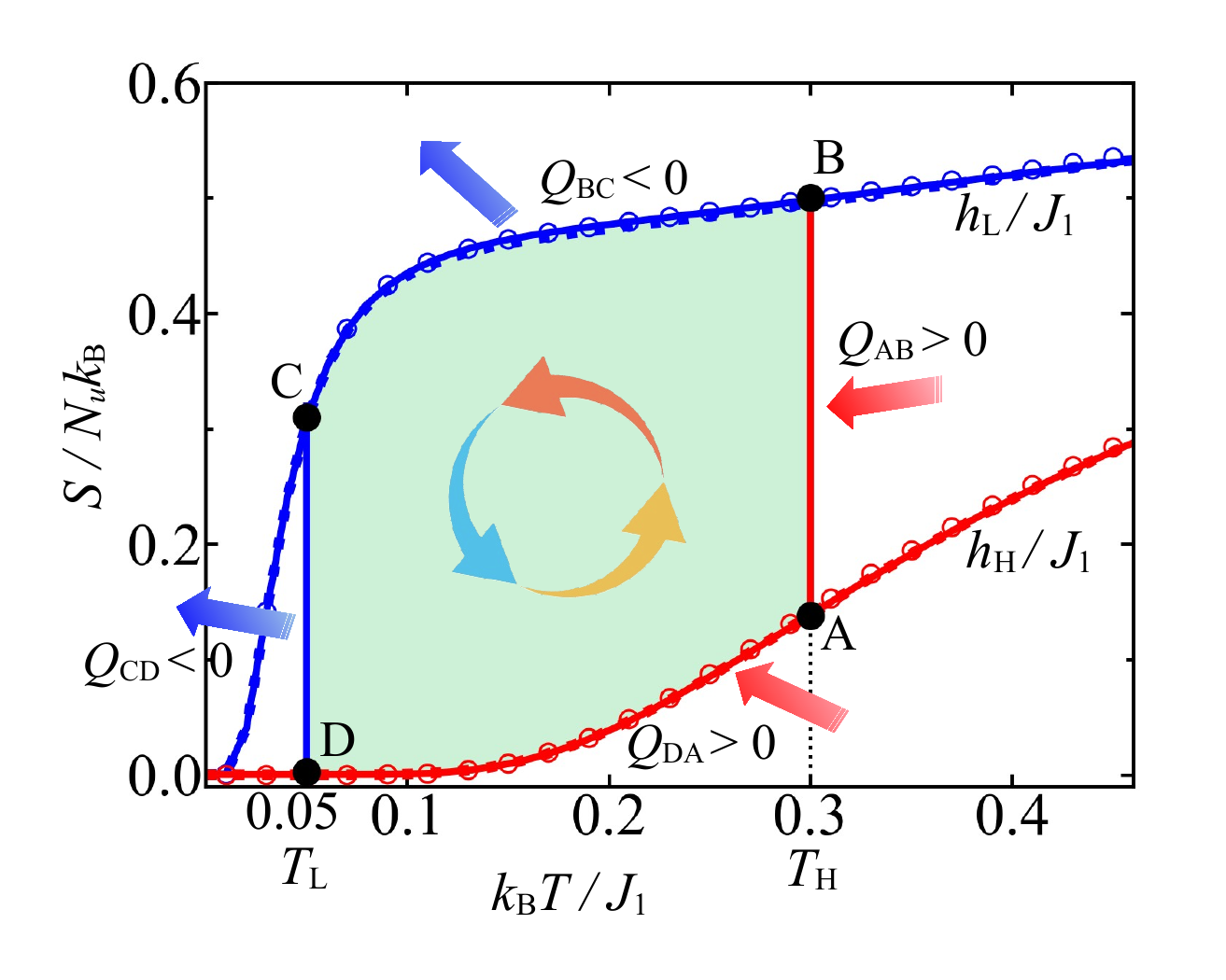}
	\caption{Quantum Stirling cycle represented in the entropy–temperature plane for the fully frustrated spin-1/2 Heisenberg four-leg tube with the fixed value of the interaction ratio $J_{2}/J_{1}=4$. Symbols denote full ED results for the number of unit cells $N_u = 4$, whereas solid and dotted lines represent the effective lattice-gas predictions for $N_u = 4$ and $N_u \to \infty$, respectively. The shaded region delimits the quantum Stirling cycle composed of two isothermal processes at $k_\text{B}T_\text{L}/J_1=0.05$ and $k_\text{B}T_\text{H}/J_1 = 0.3$, and two isofield processes at $h_\text{L}/J_1=3.9$ and $h_\text{H}/J_1=13$. Red (blue) arrows indicate heat absorbed (released) by the working medium.}
	\label{fig:Sterlingcycle}
\end{figure}

The quantum Stirling cycle consists of two isothermal and two isofield processes as illustrated in Fig.~\ref{fig:Sterlingcycle}, where we adopt the convention $Q>0$ ($Q<0$) for heat absorbed (released) by the working medium. The four thermodynamic states defining the cycle are $A=(T_H,h_H)$, $B=(T_H,h_L)$, $C=(T_L,h_L)$, and $D=(T_L,h_H)$, where $T_H>T_L$ and $h_H>h_L$. During the high-temperature isothermal process $A\to B$, the system remains in thermal equilibrium with the hot reservoir at temperature $T_H$, while the magnetic field decreases from $h_H$ to $h_L$. The working medium absorbs heat $Q_{AB}=T_H[S_B(T_H,h_L)-S_A(T_H,h_H)]>0$ from the hot reservoir. This process is subsequently followed by the isofield cooling $B\to C$ at $h=h_L$, during which the temperature decreases from $T_H$ to $T_L$. Since no work is performed at constant magnetic field, the decrease in the internal energy is entirely associated with heat released by the working medium $Q_{BC}=U_C(T_L,h_L)-U_B(T_H,h_L)<0$. During the low-temperature isothermal process $C\to D$, the system remains in thermal equilibrium with the cold reservoir at temperature $T_L$, while the magnetic field increases from $h_L$ to $h_H$. The working medium consequently releases heat $Q_{CD}=T_L[S_D(T_L,h_H)-S_C(T_L,h_L)]<0$. The cycle is completed by the isofield heating process $D\to A$ at $h=h_H$, during which the temperature increases from $T_L$ back to $T_H$ and the working medium absorbs heat while its internal energy increases $Q_{DA}=U_A(T_H,h_H)-U_D(T_L,h_H)>0$. In Fig.~\ref{fig:Sterlingcycle}, the fully frustrated spin-1/2 Heisenberg four-leg tube is considered as a working medium of this quantum Stirling cycle by specifically considering the two isothermal processes at $k_{\rm B}T_L/J_1=0.05$ and $k_{\rm B}T_H/J_1=0.3$, and the two isofield processes at $h_L/J_1=3.9$ and $h_H/J_1=13$.
Since perfect regeneration is not assumed, the heat exchanged during both isofield processes contributes to the overall heat balance of the cycle. The total heat absorbed by the working medium during one complete cycle is therefore $Q_{in}=Q_{AB}+Q_{DA}>0$, whereas the total heat released by the working medium is $Q_{out}=Q_{BC}+Q_{CD}<0$. Since the working medium returns to its initial state after one complete cycle, the total change in its internal energy vanishes. According to the first law of thermodynamics, the net work output is therefore given by:
\begin{eqnarray}
\label{Eq:work}
W = Q_{in}+Q_{out}.
\end{eqnarray} 
The heat-engine regime is realized when $W>0$, $Q_{in}>0$, and $Q_{out}<0$. The efficiency of the quantum Stirling heat engine is defined as the ratio of the net work output to the total heat absorbed by the working medium during one complete cycle:
\begin{eqnarray}\label{Eq:efficiency}
	\eta = \frac{W}{Q_{in}}=1+\frac{Q_{out}}{Q_{in}}.
\end{eqnarray}
In the absence of perfect regeneration, the efficiency depends on the heat exchanged during all four processes of the Stirling cycle.

\begin{figure}[t]
	\centering
	\includegraphics[scale=0.34,trim=0 0 0 0, clip]{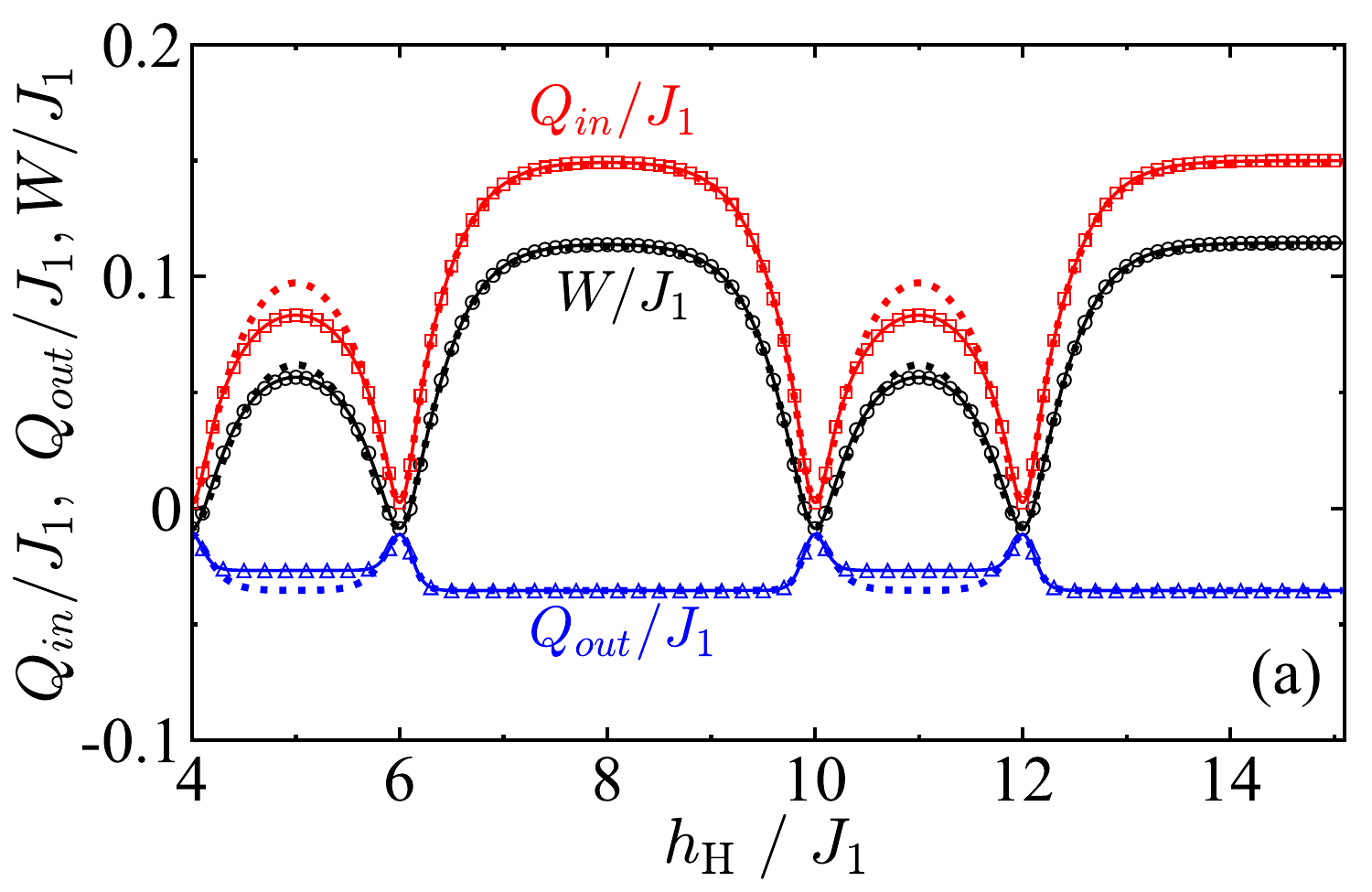}
	\includegraphics[scale=0.34,trim=0 0 0 0, clip]{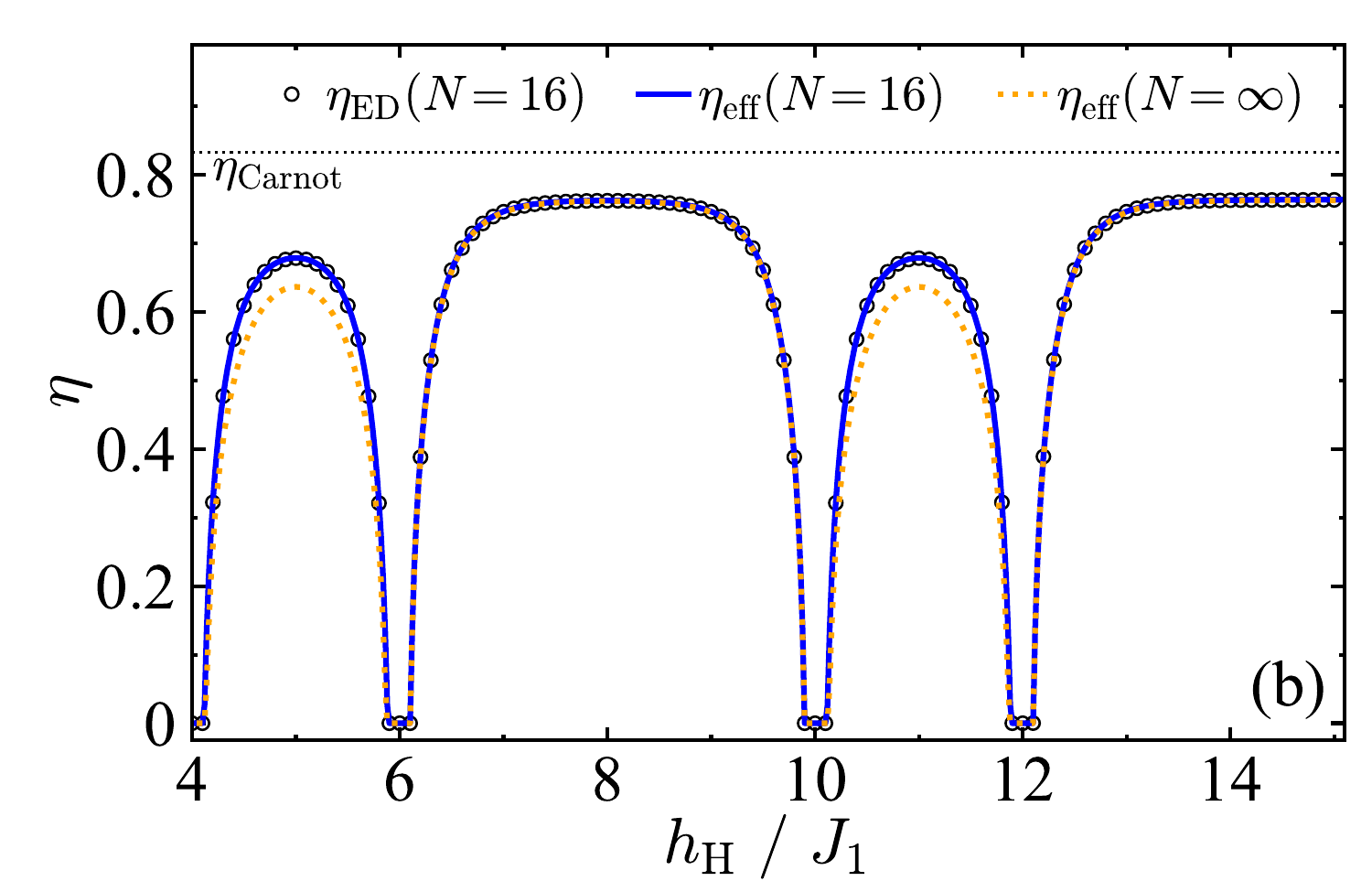}
	\caption{(a) Total absorbed heat $Q_{in}$, total released heat $Q_{out}$, and net work output $W$ of the quantum Stirling cycle involving the spin-1/2 Heisenberg four-leg tube with the ratio $J_2/J_1=4.0$ as the working substance. The cycle operates between the two temperatures $k_\text{B}T_{L}/J_1=0.05$ and $k_\text{B}T_{H}/J_1=0.3$ with the lower magnetic field fixed at $h_L/J_1=3.9$ and the upper field $h_{H}/J_1$ varied. (b) Efficiency of the quantum Stirling cycle for the same parameters as specified in panel (a). Symbols denote full ED results for $N_u=4$ unit cells, while solid and dotted lines represent the effective lattice-gas results for $N_u=4$ and $N_u\to\infty$, respectively.}
	\label{fig:WQHQL_eta}
\end{figure}

We now examine the performance of the quantum Stirling heat engine with the working medium constituted by the fully frustrated spin-1/2 Heisenberg four-leg tube, which hosts in the highly frustrated regime $J_2/J_1=4.0$ the four Wigner magnon crystals S(0), ST(0-1), T(1), and TQ(1-2). Figure \ref{fig:WQHQL_eta}(a) shows the total absorbed heat $Q_{\rm in}$, the total released heat $Q_{\rm out}$, and the net work output $W$ as a function of the upper magnetic field $h_H/J_1$, while the remaining cycle parameters are fixed at $h_L/J_1=3.9$, $k_\text{B}T_\text{L}/J_1=0.05$, and $k_\text{B}T_\text{L}/J_1=0.3$. The lower field $h_L/J_1=3.9$ is fixed to stabilize the S(0) ground state. As shown in Fig.~\ref{fig:WQHQL_eta}(a), the working substance then operates in the heat-engine regime over several intervals of the upper magnetic field $h_H$, where the conditions $Q_{\rm in}>0$, $Q_{\rm out}<0$, and $W>0$ are simultaneously satisfied. The heat exchange and work output exhibit a pronounced nonmonotonic field dependence, which closely reflects the sequence of the three remaining Wigner magnon crystals ST(0-1), T(1), and TQ(1-2), and eventually the fully polarized FM(2) phase. In particular, $Q_{\rm in}$ and $W$ are strongly suppressed in the vicinity of all field-driven phase transitions at $h_H/J_1 = 6$, $10$, and $12$, where the net work output approaches zero. The strong suppression of the work output near these phase boundaries demonstrates that proximity to a field-induced quantum phase transition does not necessarily enhance the performance for the present Stirling cycle. Instead, both quantities develop broad maxima located nearly in the middle of the field intervals corresponding to the Wigner magnon crystals ST(0-1), T(1), and TQ(1-2). Interestingly, the maxima associated with the T(1) and FM(2) phases are somewhat higher than those found within the ST(0-1) and TQ(1-2) phases. The released heat $Q_{\rm out}$ exhibits a complementary field dependence. 

The field dependence of the efficiency $\eta$ of the quantum Stirling cycle is displayed in Fig.~\ref{fig:WQHQL_eta}(b). The efficiency exhibits a sequence of the dome-like structures developed over the field intervals corresponding to the ST(0-1), T(1), TQ(1-2), and FM(2) phases, whereas pronounced minima occur in the vicinity of the respective transition fields $h_H/J_1 = 6$, $10$, and $12$. This behavior closely coincides with the field dependence of the work output and demonstrates that the most favorable operating conditions are found deep inside the stability regions of the individual ST(0-1), T(1), TQ(1-2), and FM(2) phases rather than at their phase boundaries. The efficiency $\eta$, shown in Fig. 8(b), exhibits a similar strong modulation by the underlying ground-state phase structure. It reaches high values within broad magnetic-field intervals corresponding to stable ground-state phases, whereas pronounced minima occur in the vicinity of the critical fields $h_H/J_1\simeq6$, $10$, and $12$. The local efficiency maxima reach approximately $\eta \simeq 0.68$ within the ST(0-1) and TQ(1-2) phases and $\eta \simeq 0.76$ within the T(1) and FM(2) phases, which are comparable with the two-reservoir Carnot reference value $\eta_{\rm C}=1-T_L/T_H \simeq 0.83$ presented in Fig. \ref{fig:WQHQL_eta}(b) for comparison. In contrast, the efficiency $\eta$ is strongly reduced near all field-induced phase transitions, where the work output simultaneously approaches zero.

It is also worth emphasizing the excellent agreement between the ED results and the effective lattice-gas description for the same finite system size $N_u=4$. The corresponding curves are nearly indistinguishable over the entire magnetic-field range for both the heat exchanges and the work output and only minor deviations are observed in the efficiency. The thermodynamic-limit results generally follow the same field dependence, although more noticeable finite-size effects occur within the ST(0-1) and TQ(1-2) phases, where the maxima of the efficiency are reduced from $\eta\simeq0.68$ for $N_u=4$ to $\eta\simeq0.64$ as $N_u\to\infty$. By contrast, the efficiency maxima within the T(1) and FM(2) phases are only weakly affected by the system size and remain close to $\eta\simeq0.76$. These results confirm that the effective lattice-gas model reliably captures both the characteristic field dependence and the finite-size evolution of the Stirling-engine performance.

\section{Conclusions}\label{sec:conclusions}

In this work, we have investigated the ground-state, magnetic and thermodynamic properties of the fully frustrated spin-$1/2$ Heisenberg four-leg tube in an external magnetic field. By combining complementary numerical and analytical approaches including DMRG, full ED, and an effective interacting lattice gas developed within the generalized localized-magnon theory, we obtained a comprehensive understanding of the ground-state and finite-temperature properties of this frustrated quantum spin system with locally conserved total spin on the square plaquettes.

Using large-scale DMRG calculations, we constructed the complete ground-state phase diagram in the $J_2/J_1$–$h/J_1$ parameter plane, which reveals a rich variety of quantum phases arising from the interplay between geometric frustration and the external magnetic field. These phases include the Wigner magnon crystals with character of the singlet phase S(0), the singlet-triplet phase ST(0-1), the triplet phase T(1), the triplet-quintuplet phase TQ(1-2), and the singlet-quintuplet SQ(0-2) in addition to the fully polarized FM(2) phase, the gapped Haldane phase H(2), and the two gapless Tomonaga-Luttinger quantum spin liquids TL(2) and TL(2-1). The phase diagram provides a unified picture of both discontinuous and continuous field-induced quantum phase transitions of the fully frustrated spin-1/2 Heisenberg four-leg tube.

In the highly frustrated regime, we investigated the finite-temperature properties using the effective interacting lattice-gas model constructed from bound one- and two-magnon states. Comparison with full ED results for $N=16$ spins demonstrates excellent agreement for the magnetization over the entire magnetic-field range persisting up to relatively high temperatures. The magnetic susceptibility is likewise accurately reproduced confirming that the effective lattice-gas model captures the dominant low-energy excitations associated with the discontinuous field-induced transitions between the Wigner magnon crystals. The entropy and specific heat are also well described at low and moderate temperatures, whereas noticeable deviations emerge at higher temperatures due to higher-energy excitations that are not included in the restricted localized-magnon manifold. Nevertheless, the effective lattice-gas model still reliably captures the dominant thermodynamic features of the low-energy sector at least up to moderate temperatures.

Finally, we explored the performance of a quantum Stirling heat engine using the fully frustrated spin-1/2 Heisenberg four-leg tube as the working medium. The work output and efficiency of the Stirling cycle are strongly governed by the underlying Wigner magnon ground states. Both these quantities are markedly suppressed in the vicinity of the field-induced quantum phase transitions and attain local maxima well inside the stability regions of the magnon crystals. In particular, the maximum efficiency is slightly smaller when the working substance is driven toward the two symmetry-broken ST(0-1) and TQ(1-2) phases compared with the uniform T(1) and FM(2) counterparts. The effective lattice-gas model reproduces the full ED results with excellent accuracy and additionally reveals that finite-size effects are more pronounced within the two symmetry-broken ST(0-1) and TQ(1-2) phases than within the T(1) and FM(2) phases. These results establish a direct connection between the sequence of field-induced quantum phases and the thermodynamic performance of the quantum Stirling engine.

\section*{Acknowledgments}
This work was supported by the Slovak Research and Development Agency under the contract APVV-24-0091 and by the grant of The Ministry of Education, Research, Development and Youth of the Slovak Republic under the contract VEGA 1/0298/25. H. A. Z. acknowledges the financial support provided under the postdoctoral fellowship program of P. J. \v{S}af\'arik University in Ko\v{s}ice, Slovakia.

\appendix

\section{Derivation of the one-magnon energy spectrum}
\label{app:derivation}

To derive the one-magnon energy spectrum, we consider the action of the Hamiltonian (\ref{Eq:4ll_ham}) on the basis states $|l,i\rangle$, where $l=1,2,3,4$ labels the spin within the $i$th unit cell. This leads to the following system of four mutually interconnected equations:
\begin{widetext}
	\begin{eqnarray}\label{Eq:appendix_one_magnon}
		\hat{H}|1,i\rangle &=& (E_{\rm FM} - J_2 - 4J_1 + h) |1,i\rangle + \frac{J_1}{2} \bigl( |1,i-1\rangle + |2,i-1\rangle + |3,i-1\rangle + |4,i-1\rangle \nonumber\\ 
		&& \quad + \, |1,i+1\rangle + |2,i+1\rangle + |3,i+1\rangle + |4,i+1\rangle \bigr) + \frac{J_2}{2} \bigl( |2,i\rangle + |4,i\rangle \bigr), \nonumber\\
		\hat{H}|2,i\rangle &=& (E_{\rm FM} - J_2 - 4J_1 + h) |2,i\rangle + \frac{J_1}{2} \bigl( |1,i-1\rangle + |2,i-1\rangle + |3,i-1\rangle + |4,i-1\rangle \nonumber\\ 
		&& \quad + \, |1,i+1\rangle + |2,i+1\rangle + |3,i+1\rangle + |4,i+1\rangle \bigr) + \frac{J_2}{2} \bigl( |1,i\rangle + |3,i\rangle \bigr), \nonumber\\
		\hat{H}|3,i\rangle &=& (E_{\rm FM} - J_2 - 4J_1 + h) |3,i\rangle + \frac{J_1}{2} \bigl( |1,i-1\rangle + |2,i-1\rangle + |3,i-1\rangle + |4,i-1\rangle, \nonumber \\
		&& \quad + \, |1,i+1\rangle + |2,i+1\rangle + |3,i+1\rangle + |4,i+1\rangle \bigr) + \frac{J_2}{2} \bigl( |2,i\rangle + |4,i\rangle \bigr), \nonumber\\
		\hat{H}|4,i\rangle &=& (E_{\rm FM} - J_2 - 4J_1 + h) |4,i\rangle + \frac{J_1}{2} \bigl( |1,i-1\rangle + |2,i-1\rangle + |3,i-1\rangle + |4,i-1\rangle \nonumber\\ 
		&& \quad + \, |1,i+1\rangle + |2,i+1\rangle + |3,i+1\rangle + |4,i+1\rangle \bigr) + \frac{J_2}{2} \bigl( |1,i\rangle + |3,i\rangle \bigr). 
	\end{eqnarray}
Here, $i$ labels the unit cell, while $E_{\rm FM}$ denotes the energy of the fully polarized ferromagnetic state. We introduce the relative one-magnon energy $\mathcal{E}_k=E_k-E_{\rm FM}$. Expanding the one-magnon eigenstate as $|\psi\rangle=\sum_{i,l}c_{l,i}|l,i\rangle$ ($l=1,2,3,4$) yields the following coupled equations for the probability amplitudes $c_{l,i}$ to be determined:
	\begin{eqnarray}
		(-\mathcal{E}_k - J_2 - 4J_1 + h) c_{1,i} &+&\frac{J_1}{2} \bigl( c_{1,i-1} + c_{2,i-1} + c_{3,i-1} + c_{4,i-1} + c_{1,i+1} + c_{2,i+1} + c_{3,i+1} + c_{4,i+1} \bigr) + \frac{J_2}{2} \bigl( c_{2,i} + c_{4,i} \bigr) = 0, \nonumber\\
		(-\mathcal{E}_k - J_2 - 4J_1 + h) c_{2,i} &+&\frac{J_1}{2} \bigl( c_{1,i-1} + c_{2,i-1} + c_{3,i-1} + c_{4,i-1} + c_{1,i+1} + c_{2,i+1} + c_{3,i+1} + c_{4,i+1} \bigr) + \frac{J_2}{2} \bigl( c_{1,i} + c_{3,i} \bigr) = 0, \nonumber\\
		(-\mathcal{E}_k - J_2 - 4J_1 + h) c_{3,i} &+& \frac{J_1}{2} \bigl( c_{1,i-1} + c_{2,i-1} + c_{3,i-1} + c_{4,i-1} + c_{1,i+1} + c_{2,i+1} + c_{3,i+1} + c_{4,i+1} \bigr) + \frac{J_2}{2} \bigl( c_{2,i} + c_{4,i} \bigr) = 0, \nonumber\\
		(-\mathcal{E}_k - J_2 - 4J_1 + h) c_{4,i}&+& \frac{J_1}{2} \bigl( c_{1,i-1} + c_{2,i-1} + c_{3,i-1} + c_{4,i-1} + c_{1,i+1} + c_{2,i+1} + c_{3,i+1} + c_{4,i+1} \bigr) + \frac{J_2}{2} \bigl( c_{1,i} + c_{3,i} \bigr) = 0. \nonumber
	\end{eqnarray}
	Exploiting translational invariance, we introduce for the probability amplitudes $c_{l,i}$ the plane-wave ansatz $c_{l,j}=f_l e^{ikj}$ ($l=1,2,3,4$), which transforms the system of coupled equations into momentum space taking the form:
	\begin{eqnarray}
		f_1 \Bigl[-\mathcal{E}_k - J_2 + h + J_1(\cos k - 4)\Bigr] &+& f_2 \left(J_1 \cos k + \frac{J_2}{2}\right) + f_3J_1\cos k + f_4 \left(J_1 \cos k + \frac{J_2}{2}\right) = 0, \quad\quad \nonumber \\
		f_2 \Bigl[-\mathcal{E}_k - J_2 + h + J_1(\cos k - 4)\Bigr] &+& f_1 \left(J_1 \cos k + \frac{J_2}{2}\right) + f_3 \left(J_1 \cos k + \frac{J_2}{2}\right) +  f_4 J_1\cos k = 0, \nonumber \\
		f_3 \Bigl[-\mathcal{E}_k - J_2 + h + J_1(\cos k - 4)\Bigr] &+& f_1 J_1\cos k + f_2 \left(J_1 \cos k + \frac{J_2}{2}\right) + f_4 \left(J_1 \cos k + \frac{J_2}{2}\right) = 0, \nonumber\\
		f_4 \Bigl[-\mathcal{E}_k - J_2 + h + J_1(\cos k - 4)\Bigr] &+& f_1 \left(J_1 \cos k + \frac{J_2}{2}\right) + f_2 J_1\cos k + f_3 \left(J_1 \cos k + \frac{J_2}{2}\right) = 0.
		\label{ft}
	\end{eqnarray}
\end{widetext}
The one-magnon energy spectrum is then obtained by solving the characteristic equations (\ref{ft}), which yields the four one-magnon branches given in Eq. (\ref{Eq:Epsilons}) of the main text.

\end{document}